\documentclass[cleanhead, cleanfoot,10pt,twocolumn]{asme2e}

\usepackage{empheq, amssymb}
\usepackage{lmodern}
\usepackage{amsmath}
\usepackage{amsfonts}
\usepackage{amssymb}
\usepackage{setspace}
\usepackage{bm} % bold for greek symbols
\usepackage{txfonts}
\usepackage[T1]{fontenc}
\usepackage[utf8]{inputenc} % accept different input encodings
\usepackage[dvipsnames]{xcolor} % defines colors
\usepackage[american]{babel}  
\usepackage{graphicx}
\usepackage{epstopdf}
\usepackage[noadjust]{cite} % grouped citations
\usepackage{ltxcmds}
\usepackage{mathtools}
\usepackage{subcaption}% for subfigures
\usepackage{multirow} %for tables 
\usepackage{blindtext}
\usepackage{hyperref}
\usepackage{enumitem}
\usepackage{ifthen}
\usepackage{etoolbox}
\usepackage{cancel}
\usepackage{empheq}
\usepackage{fontawesome}
\usepackage{soul}
\usepackage{flushend} 
\usepackage{textcomp}
\usepackage{lipsum}
\usepackage{nomencl}
\usepackage[displaymath,switch]{lineno}
\usepackage{fontawesome}
\usepackage{titlesec}
\usepackage{colortbl}
\usepackage{accents}
\usepackage[framemethod=TikZ]{mdframed}
\usepackage{xpatch}
\usepackage{dblfloatfix}
\usepackage{float}

\DeclarePairedDelimiter\norm{\lVert}{\rVert}

\definecolor{light-gray}{gray}{0.4}
\definecolor{box-gray}{gray}{1}

\hypersetup{
    unicode=false,          % non-Latin characters in Acrobat’s bookmarks
    pdftoolbar=true,        % show Acrobat’s toolbar?
    pdfmenubar=true,        % show Acrobat’s menu?
    pdffitwindow=false,     % window fit to page when opened
    pdfstartview={FitV},    % fits the width of the page to the window
    pdftitle={Open-Loop Control Co-Design of Semisubmersible Floating Offshore Wind Turbines using Linear Parameter-Varying Models},    % title
    pdfauthor={Athul K. Sundarrajan and Yong Hoon Lee and James T. Allison, Daniel S. Zalkind and Daniel R. Herber},     % author
    pdfsubject={},   % subject of the document
    pdfkeywords = {}, % list of keywords
    pdfnewwindow=true,      % links in new window
    colorlinks=true,
    allcolors=blue
}

\renewcommand\nomgroup[1]{%
  \item[\bfseries
  \ifstrequal{#1}{V}{ Variables}{%
  \ifstrequal{#1}{B}{ Subscripts}{%
  \ifstrequal{#1}{P}{ Notation}{%
  \ifstrequal{#1}{A}{ Acronyms}{}}}}]
}

\makenomenclature


\definecolor{block-gray}{gray}{0.95}
\makeatletter
\xpatchcmd{\endmdframed}
  {\aftergroup\endmdf@trivlist\color@endgroup}
  {\endmdf@trivlist\color@endgroup\@doendpe}
  {}{}
\makeatother

\definecolor{light-gray}{gray}{0.6}

\newcommand{\xsection}[1]{\section[#1]{\MakeUppercase{#1}}}

\definecolor{needcolor}{HTML}{C62828}

\newcommand{\openFAST}{OpenFAST}

\newcommand*\patchAmsMathEnvironmentForLineno[1]{%
  \expandafter\let\csname old#1\expandafter\endcsname\csname #1\endcsname
  \expandafter\let\csname oldend#1\expandafter\endcsname\csname end#1\endcsname
  \renewenvironment{#1}%
     {\linenomath\csname old#1\endcsname}%
     {\csname oldend#1\endcsname\endlinenomath}}%
\newcommand*\patchBothAmsMathEnvironmentsForLineno[1]{%
  \patchAmsMathEnvironmentForLineno{#1}%
  \patchAmsMathEnvironmentForLineno{#1*}}%
\AtBeginDocument{%
\patchBothAmsMathEnvironmentsForLineno{equation}%
\patchBothAmsMathEnvironmentsForLineno{align}%
\patchBothAmsMathEnvironmentsForLineno{flalign}%
\patchBothAmsMathEnvironmentsForLineno{alignat}%
\patchBothAmsMathEnvironmentsForLineno{gather}%
\patchBothAmsMathEnvironmentsForLineno{multline}%
}
\usepackage{hyperref}

\newcommand{\doi}[1]{{doi:~\href{http://doi.org/#1}{#1}}\rmFullStop}

\newcommand*{\rmFullStop}{\rmifnextchar{.}{}{}}

\makeatletter
\newcommand{\rmifnextchar}[3]{%
  \begingroup
  \ltx@LocToksA{\endgroup#2}%
  \ltx@LocToksB{\endgroup#3}%
  \ltx@ifnextchar{#1}{%
    \def\next{\the\ltx@LocToksA}%
    \afterassignment\next
    \let\scratch= %
  }{%
    \the\ltx@LocToksB
  }%
}
\makeatother
\title{Open-Loop Control Co-Design of Semisubmersible Floating Offshore Wind Turbines using Linear Parameter-Varying Models}
\author{}
\author{Athul~K.~Sundarrajan\textsuperscript{1}\thanks{Corresponding author, \texttt{\href{mailto:athul.sundarrajan@colostate.edu}{athul.sundarrajan@colostate.edu}}}, Yong~Hoon~Lee\textsuperscript{2}, James~T.~Allison\textsuperscript{3}, Daniel~S.~Zalkind\textsuperscript{4}, Daniel~R.~Herber\textsuperscript{1}}
\author{} % just trying some options here
\author{} % just trying some options here

\author{
\affiliation{
\textsuperscript{1}Department of Systems Engineering \\
Colorado State University \\
Fort Collins, CO 80523 \\
\texttt{\{\href{mailto:athul.sundarrajan@colostate.edu}{athul.sundarrajan}, \href{mailto:daniel.herber@colostate.edu}{daniel.herber}\}}@colostate.edu
}
}
\author{
\affiliation{
\textsuperscript{2}Department of Mechanical Engineering \\
The University of Memphis \\
Memphis, TN 38152 \\
\texttt{\{\href{mailto:yhlee@memphis.edu}{yhlee}\}}@memphis.edu \\
}
}

\author{
\affiliation{
\textsuperscript{3}Department of Industrial and Enterprise Systems Engineering \\
University of Illinois at Urbana-Champaign \\
Urbana, IL 61801 \\
\texttt{\{\href{mailto:jtalliso@illinois.edu}{jtalliso}\}}@illinois.edu \\
}
}
\author{\affiliation{
\textsuperscript{4}National Renewable Energy Laboratory\\
Golden, CO 80401 \\
\texttt{\{\href{mailto:daniel.zalkind@nrel.gov}{daniel.zalkind}\}}@nrel.gov
}
}

\begin{document}
\setlength{\parskip}{0pt}
\setlength{\parsep}{0pt}
\setlength{\headsep}{0pt}
\setlength{\topsep}{0pt}
\abovedisplayshortskip=3pt
\belowdisplayshortskip=3pt
\abovedisplayskip=3pt
\belowdisplayskip=3pt

\titlespacing*{\section}{0pt}{18pt plus 1pt minus 1pt}{3pt plus 0.5pt minus 0.5pt}

\titlespacing*{\subsection}{0pt}{9pt plus 1pt minus 0.5pt}{1pt plus 0.5pt minus 0.5pt}

\titlespacing*{\subsubsection}{0pt}{9pt plus 1pt minus 0.5pt}{1pt plus 0.5pt minus 0.5pt}

\maketitle
%------------------------------------------------------
\begin{abstract}\noindent
\textit{This paper discusses a framework to design elements of the plant and control systems for floating offshore wind turbines in an integrated manner using linear parameter-varying models.
Multiple linearized models derived from aeroelastic simulation software in different operating regions characterized by the incoming wind speed are combined to construct an approximate low-fidelity model of the system.
The combined model is then used to generate open-loop, optimal control trajectories as part of a nested control co-design strategy that explores the system's power production and stability using the platform pitch tilt as a proxy in the context of crucial plant and control design decisions.
The radial distance between the central and outer columns and the diameter of the outer columns of the semisubmersible platform are the plant design variables. The platform stability and power production are studied for different plant design decisions.
The effect of plant decisions on subsequent power production and stability response of the floating wind turbine is quantified in terms of the levelized cost of energy.
The results show that the inner-loop constraints and the plant design decisions affect the turbine's power and, subsequently, the cost of the system.
}
\end{abstract}

\vspace{1ex}

%----------------------------------------------------------------
% Introduction
\xsection{Introduction}\label{sec:introduction}

\begin{figure}[t]
\centering
\noindent\includegraphics[width=\columnwidth]{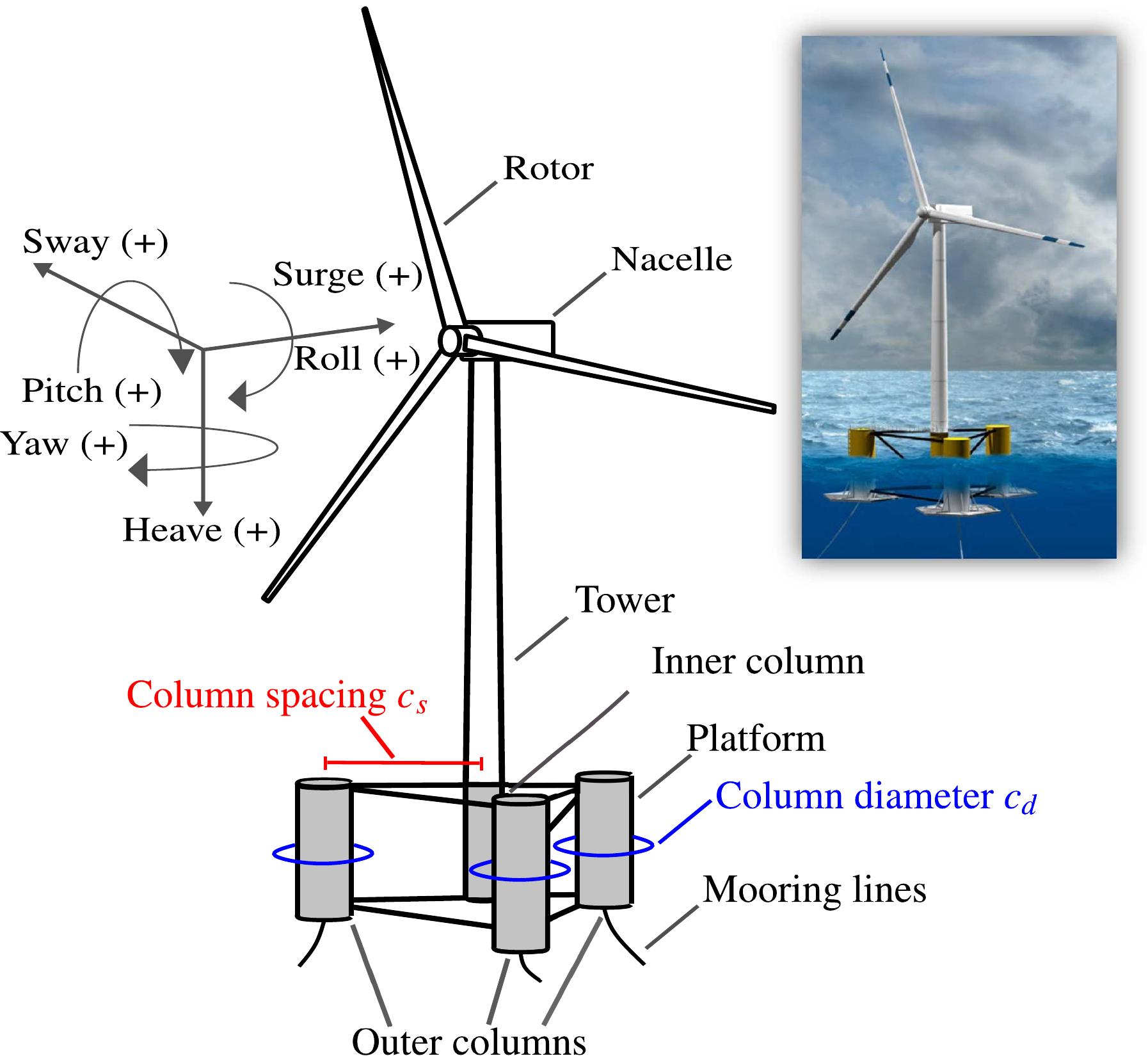}
\caption{Floating offshore wind turbine. \textit{Illustration courtesy of NREL}.}
\label{fig:fowt} %
\end{figure}

The design of floating offshore wind turbines (FOWTs) has often followed a sequential pattern, where the physical plant parameters are designed first, and a controller is then developed for a particular plant~\cite{GarciaSanz2019, Jonkman2021a, Sandner2014, Barter2020}.
However, in FOWTs, there are strong interactions between the structural and environmental dynamics and the controller.
Unfortunately, a sequential design process can produce conservative designs because it does not account for this coupling~\cite{Hegseth2020, Fleming2014}.
Optimizing both the physical plant and the controller simultaneously enables rapid identification of stable, system-level optimal results.
This integrated design approach has been studied extensively under the term control co-design (CCD)~\cite{Allison2014, Deshmukh2015, Herber2017e, Fathy2003a, Nash2019a, GarciaSanz2019}.
Recently, the importance of these integrated design approaches for energy system design has been recognized by domain experts.
References~\cite{Hegseth2020, Gaertner2020, Zalkind2019,Du2020,Muro2022} have explored the application of integrated design to offshore wind turbines.
Integrated design approaches have also found applications in design of mixed renewable/nonrenewable power generation systems~\cite{Brodrick2015,VercellinoX2}.

The primary design goal of any wind-based energy system is to capture as much power from the incoming wind while minimizing the structure's dynamic loads.
However, the overarching balance between increasing the annual energy production while minimizing the systems' building and operating costs is essential to producing economical energy solutions.
These goals are captured by the levelized cost of energy (LCOE) metric ~\cite{Gros2013}:
\begin{align}
    \label{eq:LCOE}
    \text{LCOE} = \frac{\text{Total Lifetime Cost}}{\text{Total Lifetime Energy Output}} 
\end{align}

\noindent The total lifetime costs of the FOWT system are a combination of the initial capital cost needed to build the system and the operation and maintenance costs over its lifetime. 
The capital costs are often directly linked to some of the plant design decisions~\cite{Ghigo2020, Kikuchi2019}.
 The maintenance costs and the total lifetime energy output are dependent on how the system operates and, consequently, depend on the environment and how it is controlled~\cite{Musial2019,Skaare2007}.
Recent studies have shown that advanced control strategies for offshore wind applications can increase the power extracted from the turbine and minimize the levelized cost~\cite{Ennis2018,Iori2022}.
Most conventional LCOE estimates have not incorporated detailed dynamic assessments nor the impact of novel control strategies.
In the case of highly coupled, highly constraint-sensitive systems, such as FOWTs, such considerations are imperative because of the many challenges making these systems economically viable~\cite{GarciaSanz2019}.
Additionally, overlooking the impacts of control decisions on optimal physical design is one of the pitfalls of sequential design approaches.

%------------------------------------------------------
\subsection{Plant Design for Floating Offshore Wind Turbines}
\label{sec:PlantIntro}

The plant design for a FOWT involves design decisions for several individual subsystems with considerations of stability, cost, and energy production.
The primary elements of a FOWT are the rotor, drivetrain, nacelle, tower, and support structure and are labeled in Fig.~\ref{fig:fowt}.
Stability of the FOWT about its natural equilibrium is required in all manner of wind, wave, and current excitations that the system might experience~\cite{Johannessen}. 
Reference~\cite{Hopstad2018} provides information about the current standard industry requirements of an FOWT.

An increase in the power production capacity of an FOWT increases turbine inertial and structural loads~\cite{Zalkind2019, Pao2021}.
In addition to this concern, the turbine must also withstand the forces and motions induced by the stochastic offshore environment~\cite{Zhao2020, Jonkman2011, robertson2011}.
The design of the substructure is thus a critical aspect of FOWT design.
Different substructure designs have been proposed based on ballast, buoyancy, and mooring stability concepts~\cite{Thiagarajan2014, butterfield2007, Subbulakshmi2022}.
The focus of this study will be the semisubmersible platform technology, which has been shown to have potential benefits over other alternatives in terms of stability, transportation, and ease of assembly~\cite{Thiagarajan2014,Musial2007}. 
In this study, the plant variables under consideration are the distance between the central column and the outer columns, also called column spacing~$(c_s)$, and the diameter of the outer columns~$(c_d)$, as they directly affect the geometry and cost of the platform:
\begin{align}
\label{eq:plant_var}
\bm{x}_p = \begin{bmatrix}c_s & c_d \end{bmatrix}^T
\end{align}

Generally, increasing the size of the support structure will make the FOWT more stable, but this would also raise the capital and other associated costs. 
Therefore, it is essential to optimize the system for cost while ensuring stability~\cite{DNV}.
The effect of other variables, such as ballast volume and mooring parameters, could also be explored.
As the development cycle progresses, additional practical considerations may also be incorporated into the plant design, like assembly costs and procedures, maintenance costs, and ease of transportation.

%------------------------------------------------------
\subsection{Control Design for Floating Offshore Wind Turbines}\label{sec:control-design}

The control system for an FOWT is instrumental in achieving the design goals stated in the previous sections.
The power generated by an FOWT and the physical loads on its structure are heavily dependent on the loading conditions induced by the wind, waves, and currents.
Operating the system in such a way so that it can remain stable while producing maximal power is the primary goal of the FOWT control system.
Similar to the control of land-based wind turbines, the control strategy selected depends heavily on the system's input excitations because these inputs produce the dynamical responses we seek to optimize.

\begin{figure}[t]
\centering
\includegraphics[width = \columnwidth]{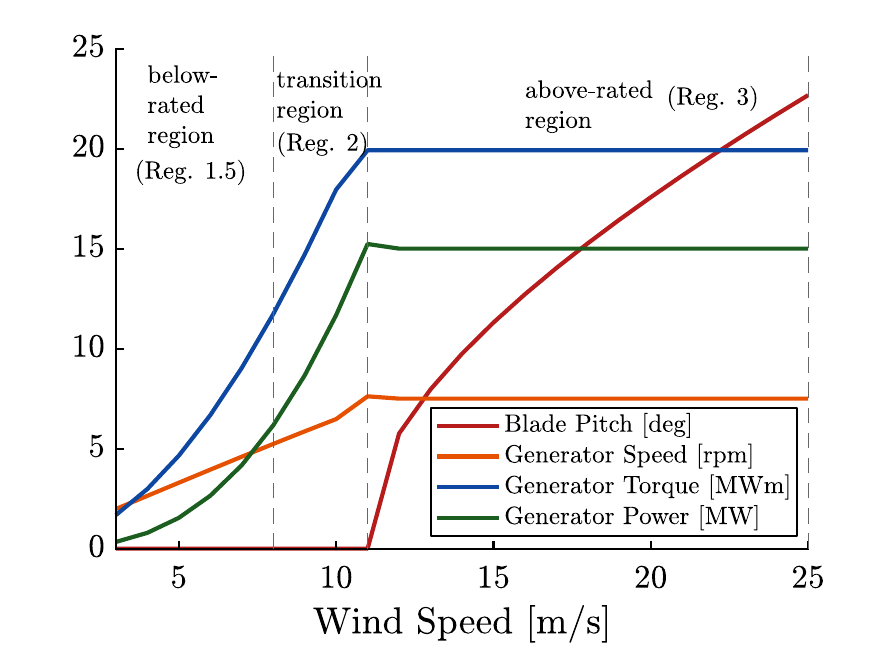}
\caption{Stationary operating points for IEA-15-MW turbine.}
\label{fig:control profiles} %
\end{figure}

% new paragraph
The primary mode of control for any wind turbine depends heavily on the wind, so specific operating regions are often defined based on the wind speed~\cite{Pao2009, Moriarty2009}.
Typically, there are three wind speed-based regions of interest, visualized in Fig.~\ref{fig:control profiles}.
At lower, below-rated wind speeds, the goal is to use the generator torque to change the generator speed that tracks the optimum power coefficient.
In the above-rated wind speed (Reg. 3), the turbine is designed to operate at its maximum power level. 
In between these regions, there is a transition behavior, and, above the cut-out wind speed, the system is shut down because there can be permanent structural damage.

The two primary control inputs for wind turbines are the pitch angle of the turbine blades (commonly called blade pitch) and the torque produced by the generator.
In below-rated wind speeds, varying the generator torque is the primary mode of control of the turbine~\cite{Gaertner2020,Abbas2021}.
Above rated wind speeds, the generator torque is held constant and the blade pitch is varied to regulate the generator speed and power to their rated values.

%------------------------------------------------------
\subsection{Modeling Considerations}
\label{sec:modeling-considerations}

It is often necessary to conduct early-stage design studies to understand the desired fundamental system properties and behaviors that inform critical decisions that need to be made as the system of interest is realized.
The use of high-fidelity modeling tools and methods in early-stage design studies is not always needed to achieve the desired design insights and can be prohibitive due to their complexity and computational expense~\cite{Herber2019b}.
In the context of optimization-based studies, depending on the parameterization of the given turbine and platform model, the resulting design space could be broad and complex~\cite{Chen2021, Hall2014}.

% new paragraph
To facilitate these design and control (both closed- and open-loop) studies, it is common to develop reduced or lower-order models that capture just the system's essential physics.
Results from these reduced-order models are validated against the results from high-fidelity tools to understand their veracity in studying the system's behavior.
In some cases, these models are then linearized around predetermined set-point values in distinct operating regions.
These linearized models are then either used to understand the system dynamics and design controllers in these operating regions or to develop frequency domain models that enable faster model evaluation~\cite{Hegseth2020, Lemmer2018, Dou2020, Faraggiana2022}.
Some recent platform design studies have utilized these linearized models and optimization-based approaches to identify the optimal design~\cite{Dou2020, Hegseth2020, Faraggiana2022}.

However, there are some challenges in developing these lower-order models.
For example, it can be complicated because this process requires extensive subject knowledge of FOWTs and the associated physics/engineering disciplines.
Additionally, the lower-order models are developed to study a specific aspect of the system's behavior (e.g.,~the floating structure response, controller response, and aerodynamic wake).
As such, the results from these models cannot be easily generalized to obtain system-level insights.
The highly coupled nature of an FOWT can create further complications in modeling the system accurately~\cite{Lemmer2018, Smilden2016, Hegseth2020, Pereira2014}.

% new paragraph
One way to mitigate these challenges is by using linearized models obtained directly from high-fidelity modeling tools (e.g.,~computational fluid dynamics, blade element momentum theory)~\cite{Jonkman2016, Jonkman2018}.
These models are obtained by linearizing the nonlinear system around specific operating points, often stationary points where the system exhibits static behavior.
A linear time-invariant state-space dynamic model about the static operating point $(\bm{\xi}_o,\bm{u}_o)$ typically has the following form:
\begin{subequations}
\label{eq:linearized-lti}
\begin{align}
\frac{d\bm{\xi}_{\Delta}(t)}{dt} &= \bm{A}_{o} \bm{\xi}_{\Delta}(t) + \bm{B}_o \bm{u}_{\Delta}(t)
\\
\bm{y}(t) &= \bm{C}_o \bm{\xi}_{\Delta}(t) + \bm{D}_o \bm{u}_{\Delta}(t) + \bm{g}_o
\end{align}
\end{subequations}

\noindent where $t$ is time, $\bm{\xi}_{\Delta}(t)$ are the relative states related to the original states $\bm{\xi}$ with $\bm{\xi}(t) = \bm{\xi}_{\Delta}(t) + \bm{\xi}_o$, $\bm{u}_{\Delta}(t)$ are the relative inputs related to the original inputs $\bm{u}$ with $\bm{u}(t) = \bm{u}_{\Delta}(t) + \bm{u}_o$, $\bm{y}(t)$ are the outputs, and the matrices $(\bm{A}_{o},\bm{B}_{o},\bm{C}_{o},\bm{D}_{o},\bm{g}_{o})$ are associated with the linearization process.

% new paragraph
A significant drawback with any kind of linearized model is that its accuracy in capturing the system's dynamic response diminishes quickly as the system's behavior moves away from the initial operating point.
Thus, it becomes difficult to work with many diverse design load cases where the wind speed continuously varies.
Some studies that have used linearized models have leveraged them in gain scheduling approaches to account for nonlinearities.
However, this approach does not guarantee stability and performance for all possible values of the wind speed~\cite{Wingerden2009}.

In this work, we will discuss the use of linear parameter-varying (LPV) models to help overcome the drawbacks of distinct linear models~\cite{Bianchi2005, Wingerden2009}.
These LPV models show good accuracy when capturing the original nonlinear dynamics and can be used to generate open-loop optimal control trajectories. 
LPV models have also been used to investigate various closed-loop control solutions for wind turbines~\cite{Wingerden2009, Lescher, Martin2017}.
However, these studies have not explored the use of continuous LPV models to approximate the nonlinear system response or its efficient application in early-stage design studies with open-loop optimal control.

%------------------------------------------------------
\subsection{Integrated Design with Control Co-Design}
\label{sec:Nested}

CCD is an integrated design paradigm that enables simultaneous design optimization of the plant and control systems~\cite{Fathy2003a, Herber2018, SundarrajanX1, Azad2019a}.
The CCD approach provides a rigorous framework that can naturally handle the coupling between the plant and control drivers present in FOWTs. 
A common mathematically equivalent way to decompose a CCD problem is with the nested formulation as a bilevel optimization~\cite{Herber2018, SundarrajanX1}.
The coordination approach defines a first-level, outer-loop problem that optimizes the plant design with information on the best possible performance from the second-level, inner-loop problem that optimizes the dynamics and control for a given plant design (and is sometimes called the control subproblem).
In other words, the outer loop generates candidate plant designs, denoted by $\bm{x}_p^{\dagger}$; this candidate is then passed to the inner loop.
The inner loop then produces an optimal control solution, $\bm{u}$, and system dynamic states, $\bm{\xi}$, for this candidate plant design.

% new paragraph
There are certain advantages to using the nested CCD approach (many are discussed in~\cite{SundarrajanX1}), especially for problems where the inner loop is a linear-quadratic dynamic optimization (LQDO) problem.
LQDO problems are characterized by quadratic objectives, linear dynamic systems, general linear constraints, and open-loop control~\cite{Herber2019b, SundarrajanX1}.
Such problems can be solved efficiently and accurately using quadratic programming methods~\cite{Herber2020d}.
Additionally, nested CCD is often necessary when black-box models of the dynamics are used (as will be the case in this work)~\cite{Deshmukh2015, Deshmukh2017}.
In this article, we demonstrate the use of the nested CCD method in the design of FOWT with the primary goal of minimizing the LCOE.
Factors such as power generation and the dynamic stability of the system are incorporated as inner-loop objectives and constraints, respectively.

% subsubsection
\subsection{Open-Loop vs.~Closed-Loop Control}
As is true in many domains, various closed-loop control strategies have been used for wind turbine control.
While these strategies are providing many practical control solutions, their use in early-stage design studies can limit exploration because a control architecture must be assumed, potentially limiting our understanding of various trade-offs that can inform better wind turbine designs~\cite{Deshmukh2015a}.
Since open-loop optimal control (OLOC) does not assume a particular control architecture, it can help identify the maximum achievable performance limits and provide critical insights into the optimal system dynamics and controller behavior in early-stage design studies~\cite{SundarrajanX1}.

In the study of many controlled dynamic systems, simulation or shooting-based approaches have been used where a simulation is performed given a controller (either of the closed-loop or open-loop variety), and its result (e.g., power generated) is used to assess the performance of the proposed control strategy.
While implementing a shooting approach is relatively straightforward, there are several challenges when combined with OLOC, such as various efficiency and convergence issues~\cite{Betts2010, Biegler2010}. 
Therefore, we use the direct-transcription~(DT) method to solve the OLOC problem, which discretizes the states and controls, resulting in a large, sparse optimization problem~\cite{Betts2010, Biegler2010, Allison2014}.

%-----------------------------------------------------------------------
\subsection{Use of OpenFAST and WEIS Models}

Wind energy with integrated servo-control (WEIS) is an open-source project that is developed by the National Renewable Energy Laboratory (NREL) and partners that will allow users to perform CCD of FOWT systems~\cite{WEIS, Jonkman2021a}.
The WEIS toolbox is built on \openFAST~\cite{openFAST}, another open-source toolbox developed by NREL, that generates a full-system dynamic response of FOWTs under wind, wave, and current excitations.
The \openFAST{} tool is built on independent modules that capture the important physical phenomena of the different FOWT subsystems and couplings between them.
There are different modules to capture the effects of aerodynamics, hydrodynamics, servodynamics, and mooring dynamics.
A variety of plant design decisions can be explored within these tools as well~\cite{Jonkman2021a}.
The Wind-Plant Integrated System \& Engineering Model (WISDEM\textsuperscript{\textregistered}), also part of WEIS, is used to compute the platform geometry, mass, and cost.

% new paragraph
In this work, the dynamic models of FOWTs will be generated using the linearization capabilities of the WEIS/\openFAST{} tools, with the original nonlinear dynamics simulation capabilities being used for validation of the results.
A detailed discussion regarding the linearization capabilities of \openFAST{} and the entire tool can be found in~\cite{Jonkman2018, Jonkman2016, openFAST, WEIS}.
Wind speed is used to select the state and control operating points for this linearized model.

The remainder of the paper is organized as follows:
Sections~\ref{sec:LPV} and \ref{sec:IEA15MW} define LPV modeling theory and validate the specific LPV models used in this work, respectively.
Section~\ref{sec:OCP} formulates the CCD problem using the LPV dynamic model.
Section~\ref{sec:results} presents the results of several studies conducted to better understand the impact of control and plant decisions on the LCOE objective.
Section~\ref{sec:conclusion} summarizes the results and provides future steps for this work.
%---------------------------------------------------------
% LPV modeling
\xsection{Linear Parameter-Varying Models}\label{sec:LPV}

As mentioned in Sec.~\ref{sec:modeling-considerations}, linearized models like the one defined in Eq.~(\ref{eq:linearized-lti}) can accurately describe the system's behavior for small perturbations about the operating point from which they were derived.
For the design and optimization activities of an FOWT system, it is essential to understand the system behavior over multiple input excitations.
While there are additional drivers for modeling variations, the primary one in wind energy systems, including FOWTs, is the wind speed in the direction of the turbine-blade system.
Under different wind conditions, the stationary operating points for the FOWT system vary greatly, as do the matrices defining the dynamic model in Eq.~(\ref{eq:linearized-lti}). 
Therefore, we will consider models dependent on this important parameter, which will be useful in OLOC CCD studies.

%-----------------------------------------------------------
\subsection{Linear Parameter-Varying Model Derivation}\label{sec:LPVModelling}

LPV models are a special case of linear time-varying (LTV) systems where the system matrices are continuous and are a function of a set of parameters~\cite{Bianchi2005, Martin2017}.
Here, we will consider the single parameter case where the parameter $w$ indicates the current wind speed value.
Now consider the following nonlinear parameter-dependent model $\Sigma$:
% \begin{subequations}
\begin{align}
\label{eq:nonlin}
\Sigma =
\begin{cases}
\displaystyle \frac{d\bm{\xi}}{dt} = \bm{f}(\bm{\xi},\bm{u},w) \\
\hspace{0.095in} \bm{y} = \bm{g}(\bm{\xi},\bm{u},w)
\end{cases}
\end{align}
% \end{subequations}

Our goal is to linearize this model about the $w$-varying operating point functions $(\bm{\xi}_o(w),\bm{u}_o(w))$ where stationary or steady-state models characterize their values:
\begin{align}
\label{eq:stationary-condition}
\bm{f}(\bm{\xi}_o(w),\bm{u}_o(w),w) = \bm{0},\ \forall w \in [w_{\min},w_{\max}]
\end{align}

\noindent where $w_{\min}$ is the minimum parameter value considered, and $w_{\max}$ is the maximum parameter value considered.

% new paragraph
Now the relationship between the linearization states and the original states depends on the parameter $w$:
\begin{align}
\label{eq:operatingpoint}
\bm{\xi}(t) &= \bm{\xi}_{\Delta}(t) + \bm{\xi}_o(w), \quad \bm{u}(t) = \bm{u}_{\Delta}(t) + \bm{u}_o(w)
\end{align}

\noindent Assuming that $w$ is time varying, the time derivative relationship of the states is:
\begin{subequations}
\label{eq:op-derivative}
\begin{align}
\frac{d\bm{\xi}}{dt} &= \frac{d\bm{\xi}_{\Delta}}{dt} + \frac{d}{dt}\bm{\xi}_o(w(t)) \\
&= \frac{d\bm{\xi}_{\Delta}}{dt} + \frac{\partial \bm{\xi}_o}{\partial w} \frac{d w}{d t}
\end{align}
\end{subequations}

Now we use the  following notation for the derivatives of the nonlinear model:
\begin{subequations}
\begin{gather*}
\bm{A}(w) \coloneqq {J}^{\bm{f}}_{\bm{\xi}}\left(\bm{\xi}_o(w),\bm{u}_o(w),w\right), \ \bm{B}(w) \coloneqq {J}^{\bm{f}}_{\bm{u}}\left(\bm{\xi}_o(w),\bm{u}_o(w),w\right) \\
\bm{C}(w) \coloneqq {J}^{\bm{g}}_{\bm{\xi}}\left(\bm{\xi}_o(w),\bm{u}_o(w),w\right), \ \bm{D}(w) \coloneqq {J}^{\bm{g}}_{\bm{u}}\left(\bm{\xi}_o(w),\bm{u}_o(w),w\right)
\end{gather*}
\end{subequations}

\noindent where $\bm{J}^{\bm{f}}_{\bm{x}}$ denotes the Jacobian of $\bm{f}$ with respect to $\bm{x}$, and the values of these functions are dependent on the operating points and are denoted as:
\begin{align*}
\bm{f}(w) \coloneqq \bm{f}\left(\bm{\xi}_o(w),\bm{u}_o(w),w\right), \
\bm{g}(w) \coloneqq \bm{g}\left(\bm{\xi}_o(w),\bm{u}_o(w),w\right)
\end{align*}

\noindent With this derivative relationship in Eq.~(\ref{eq:op-derivative}) and the notation above, the nonlinear system $\Sigma$ in Eq.~(\ref{eq:nonlin}) is linearized about $(\bm{\xi}_o(w),\bm{u}_o(w))$ yielding the following LPV system:
% \begin{subequations}
\begin{align}
\label{eq:sigma-w}
\Sigma_w =
\begin{cases}
\displaystyle\frac{d\bm{\xi}_{\Delta}}{dt} = \cancelto{0}{\bm{f}(w)}
+ \bm{A}(w) \bm{\xi}_{\Delta}
+ \bm{B}(w) \bm{u}_{\Delta}
- \displaystyle\frac{\partial \bm{\xi}_o(w)}{\partial w} \frac{d w}{d t}
\\
\hspace{0.163in} \bm{y} = \bm{g}(w)
+ \bm{C}(w) \bm{\xi}_{\Delta}
+ \bm{D}(w) \bm{u}_{\Delta}
\end{cases}
\end{align}
% \end{subequations}

If only a single time-invariant value of the parameter denoted $w_o$ is considered, then we have the following system:
\begin{subequations}
\begin{align}
\frac{d\bm{\xi}_{\Delta}}{dt} &= \bm{A}(w_o) \bm{\xi}_{\Delta}
+ \bm{B}(w_o) \bm{u}_{\Delta}
- \frac{\partial \bm{\xi}_o(w_o)}{\partial w} \cancelto{0}{\frac{d w}{d t}}
\\
\bm{y} &= \bm{g}(w_o)
+ \bm{C}(w_o) \bm{\xi}_{\Delta}
+ \bm{D}(w_o) \bm{u}_{\Delta}
\end{align}
\end{subequations}

\noindent which gives us:
% \begin{subequations}
\label{eq:singleOP}
\begin{align}
\Sigma_o =
\begin{cases}
\displaystyle\frac{d\bm{\xi}_{\Delta}}{dt} = \bm{A}(w_o) \bm{\xi}_{\Delta}
+ \bm{B}(w_o) \bm{u}_{\Delta}
\\
\hspace{0.163in} \bm{y} = \bm{g}(w_o)
+ \bm{C}(w_o) \bm{\xi}_{\Delta}
+ \bm{D}(w_o) \bm{u}_{\Delta}
\end{cases}
\end{align}
% \end{subequations}
\noindent which is the same LTI system defined in Eq.~(\ref{eq:linearized-lti}) for a single operating point characterized by the parameter $w_o$.

%-----------------------------------------------------------
\subsection{Construction Using Multiple Linearized Models}\label{sec:LPVconstruction}

The system $\Sigma_w$ with continuous dependence on the parameter $w$ generally will not be directly available because linearized models are often realized through numerical methods for specific operating points (i.e.,~$\Sigma_o$). 
Therefore, it may be necessary to construct $\Sigma_w$ from a finite strategic set of $\Sigma_o$ models.
To accomplish this goal, the matrix entries of $\Sigma_w$ are determined by element-wise matrix interpolation from a set of given denoted $\bm{\Omega} = \{ \Sigma_{o1}, \Sigma_{o2}, \cdots, \Sigma_{on} \}$, each created using the parameters values $\bm{W} = [w_1, w_2, \cdots, w_n]$.
The selected interpolation scheme was piecewise cubic Hermite interpolating polynomials.
Derivatives of the polynomial interpolating function are directly computed when needed.

% new paragraph
There are several properties to consider to ensure such an interpolation scheme has a reasonable chance of meaningfully capturing the nonlinear dynamics, including:
\begin{enumerate}[label=(P\arabic*)]
\item The structure of the states, inputs, and outputs are unchanging for all considered $\Sigma_o$.

\item The sparsity patterns (nonzero entries in the system matrices) are generally similar between analogous matrices.\label{test:matrixsparsity}

\item The stationary condition in Eq.~(\ref{eq:stationary-condition}) holds for the given interpolation scheme and $\bm{W}$ (i.e.,~$(\bm{\xi}_o(w), \bm{u}_o(w))$) can be found through interpolation such that the condition holds.

\item The element-wise relationships between different matrices can be reasonably interpolated using a selected $\bm{W}$; however, this is hard to quantify because errors in these coefficients might not result in large errors in the key outputs.\label{test:MatrixIntp}

\item At various validation points not in $\bm{W}$, the error between the actual linearized system at $w_o$ and the interpolated system $\Sigma_w$, quantified by the $H_\infty$ norm, is below a tolerance $\epsilon$:
\begin{align}\label{eq:Hinf}
\norm*{\bm{G}_o(s) - \bm{G}_w(s)}_{H_\infty} \leq \epsilon
\end{align}

where $\bm{G}_o(s)$ and $\bm{G}_w(s)$ are the transfer function matrices for $\Sigma_o$ and $\Sigma_w$, respectively.
This error metric better captures the input/output error between the interpolated and original systems.
\label{test:tf}

\item Time-domain~simulations between the nonlinear $\Sigma$ and LPV $\Sigma_w$ should be similar.\label{test:TDSimulation}

\end{enumerate}

At this time, the selection of $\bm{W}$ was informed by expert intuition and figures such as Fig.~\ref{fig:control profiles} that characterize the different regions of operation and their transition points.
Future work will consider automated sampling strategies that try to optimally sample points for constructing an accurate LPV using the condition in Eq.~(\ref{eq:Hinf}).
%_-------------------------------------------------------------
% Validation
\section{LPV Model Validation for IEA-15 MW Turbine}\label{sec:IEA15MW}

\begin{figure*}[t]
\centering
\includegraphics[scale = 0.5]{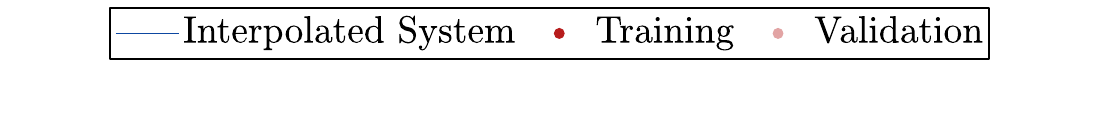}\\
\begin{subfigure}[t]{0.32\textwidth}
\centering
\noindent\includegraphics[scale=0.35]{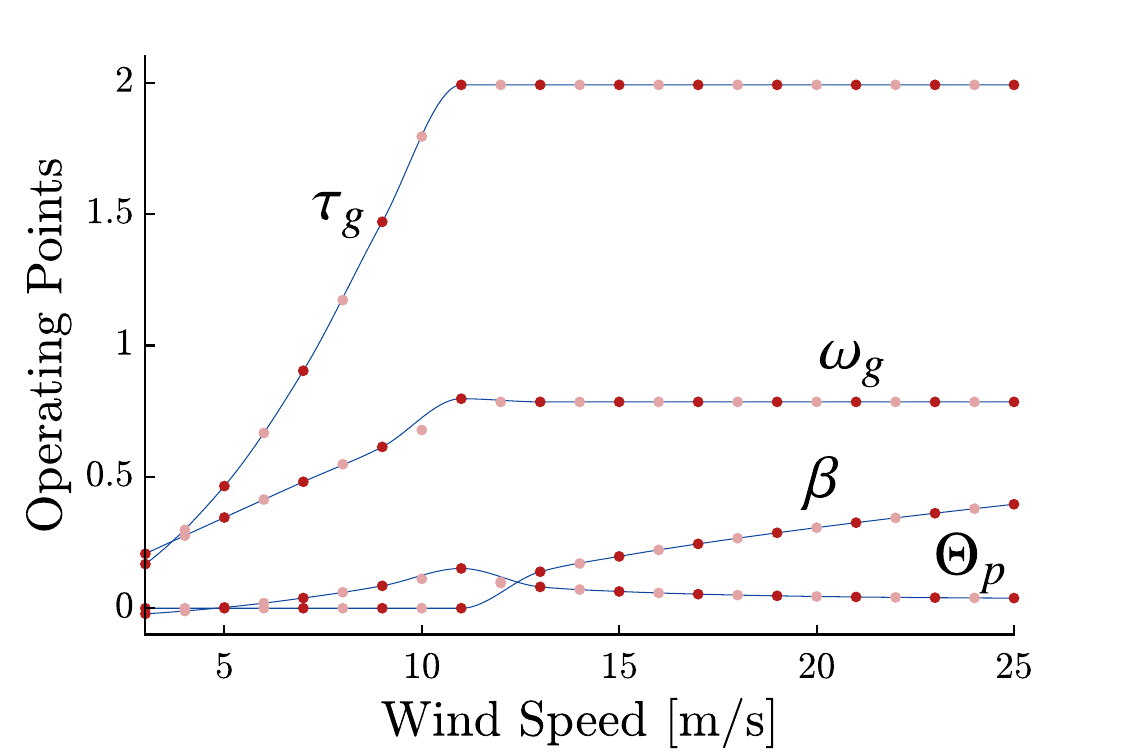}
\caption{Select stationary operating point values.}
\label{fig:OP}
\end{subfigure}%
\hspace{0.015\textwidth}%
\begin{subfigure}[t]{0.32\textwidth}
\centering
\noindent\includegraphics[scale=0.35]{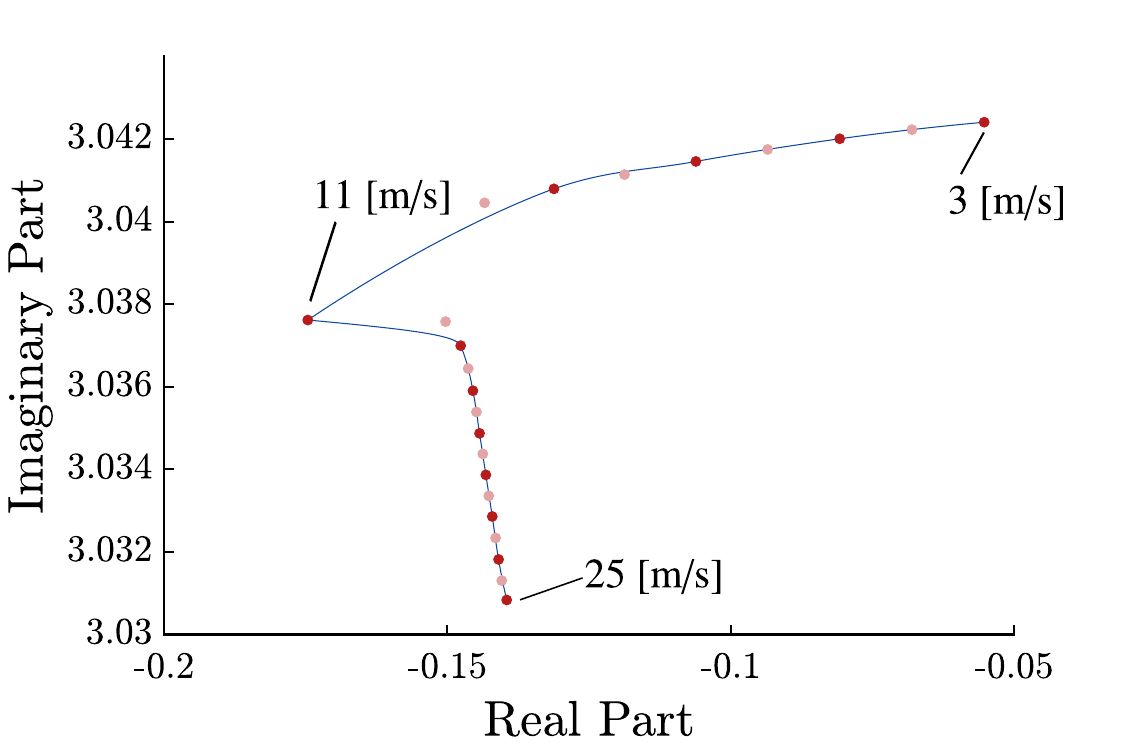}
\caption{One eigenvalue of $\bm{A}(w)$.}
\label{fig:IntMatA} 
\end{subfigure}%
\hspace{0.015\textwidth}%
\begin{subfigure}[t]{0.32\textwidth}
\centering
\noindent\includegraphics[scale=0.35]{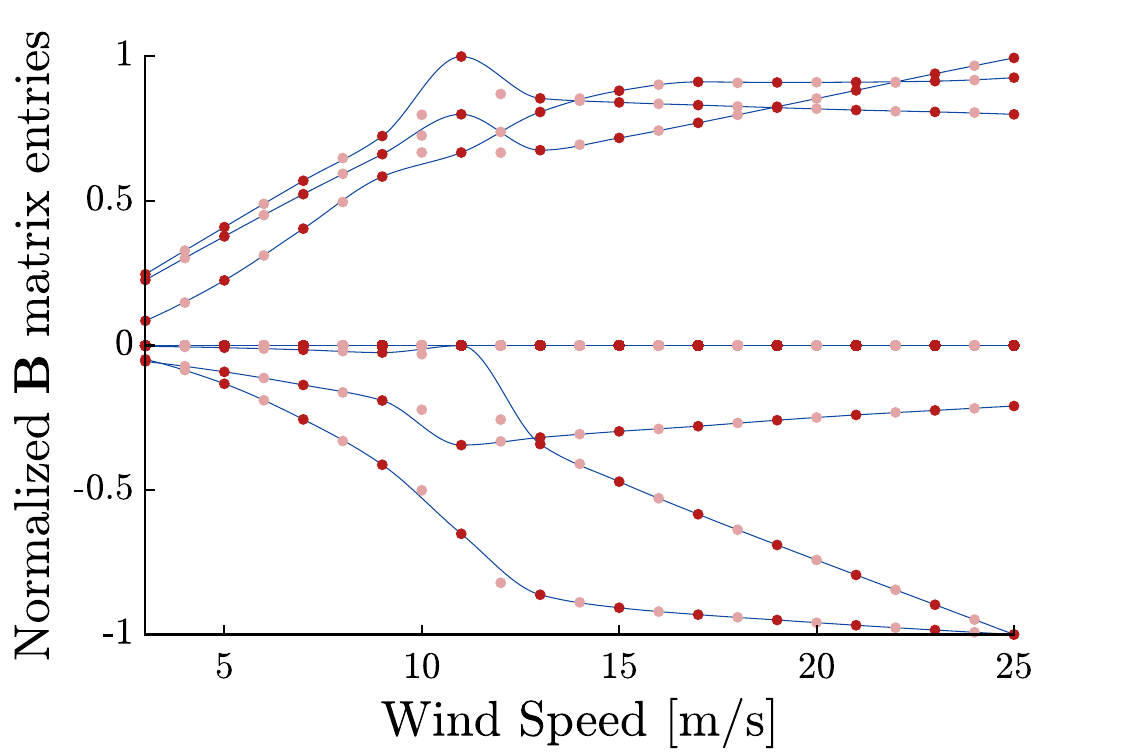}
\caption{Interpolated matrix entries of $\bm{B}(w)$.}
\label{fig:IntMatB}
\end{subfigure}%
%\noindent \includegraphics[width = \textwidth]{figures/S3/figEIG/eignevalues_combined.pdf}
\caption{Select stationary points, eigenvalues, and input matrix entries for $\Sigma_w$ for the IEA 15-MW wind turbine.}
\label{fig:MatValidation}
\end{figure*}

%\xneed{\blindtext}
The International Energy Agency (IEA) 15-MW offshore wind turbine is a reference turbine model jointly developed by NREL and Danish Technical University (DTU)~\cite{Gaertner2020, IEA15MW}, visualized in Fig.~\ref{fig:fowt}.
The turbine is supported by a floating semisubmersible platform and a chain catenary mooring system. The details of the support structure are available in~\cite{Allen2020}.
This is the system under consideration in this work.

%-----------------------------------------------------------------
There are five states, namely the platform pitch $\Theta_p$, the first time derivative of platform pitch $\Dot{\Theta}_p$, the tower-top fore-aft displacement $\delta_T$, first time derivative of the tower-top fore-aft displacement $\Dot{\delta}_T$, and the generator speed $\omega_g$.
The order of the states is as follows:
\begin{gather}
    \label{eq:states}
    \bm{\xi}(t) = \begin{bmatrix}
        \Dot{\Theta}_p & \Theta_p & \Dot{\delta}_T &\delta_T & \omega_g
    \end{bmatrix}^T
\end{gather}
In the rest of this article, we restrict our focus to two key states, namely the generator speed $\omega_g$ and platform pitch $\Theta_p$, but all are included in the LPV models.
In its current form, the model is excited by wind inputs only; wave and current disturbances are not considered.
Correspondingly, the total inputs to the system are the wind speed $w$, the generator torque $\tau_g$, and the blade pitch $\beta$:

\begin{gather}
\label{eq:controls}
\bm{u}(t) = \begin{bmatrix} \tau_g & \beta \end{bmatrix}^T % \\
\end{gather}

For the considered system, \openFAST{} can provide accurate simulations of the system's nonlinear dynamics (i.e.,~the outputs of $\Sigma$).
However, due to the concerns expressed in previous sections, an LPV model is considered a less computationally expensive and structured alternative to these expensive simulations.
The natural choice for the parameter needed to construct the LPV model $\Sigma_w$ is the wind speed.
The operating region of a wind turbine is between the cut-in wind speed ($w_{\min} = 3$ [m/s] in this study) and the cut-out wind speed ($w_{\max} = 25$ [m/s]).
To understand the accuracy of the LPV modeling approach for this system, several validation studies were carried out.

\subsection{State-Space Model Comparisons}
\label{sec:SS-model-val}

\begin{figure*}[t]
\centering
\begin{subfigure}[t]{0.32\textwidth}
\centering
\noindent\includegraphics[scale=0.35]{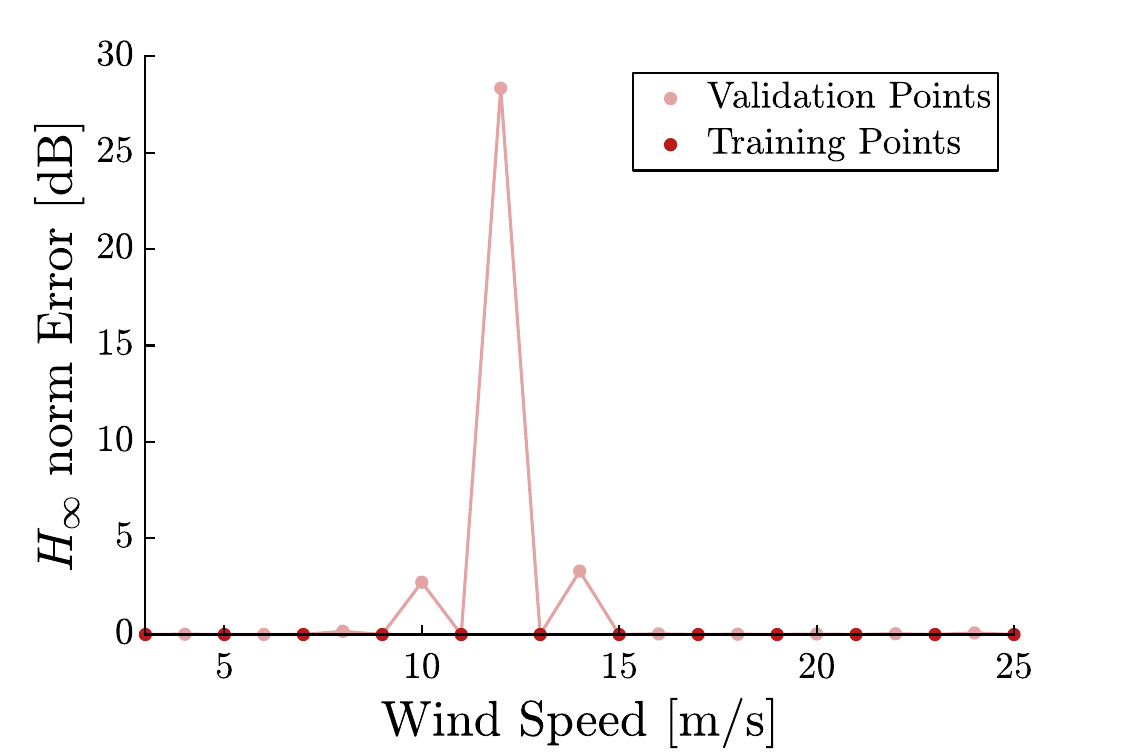}
\caption{$H_\infty$ error using validation points.}
\label{fig:Hinf}
\end{subfigure}%
\hspace{0.015\textwidth}%
\begin{subfigure}[t]{0.32\textwidth}
\centering
\noindent\includegraphics[scale=0.35]{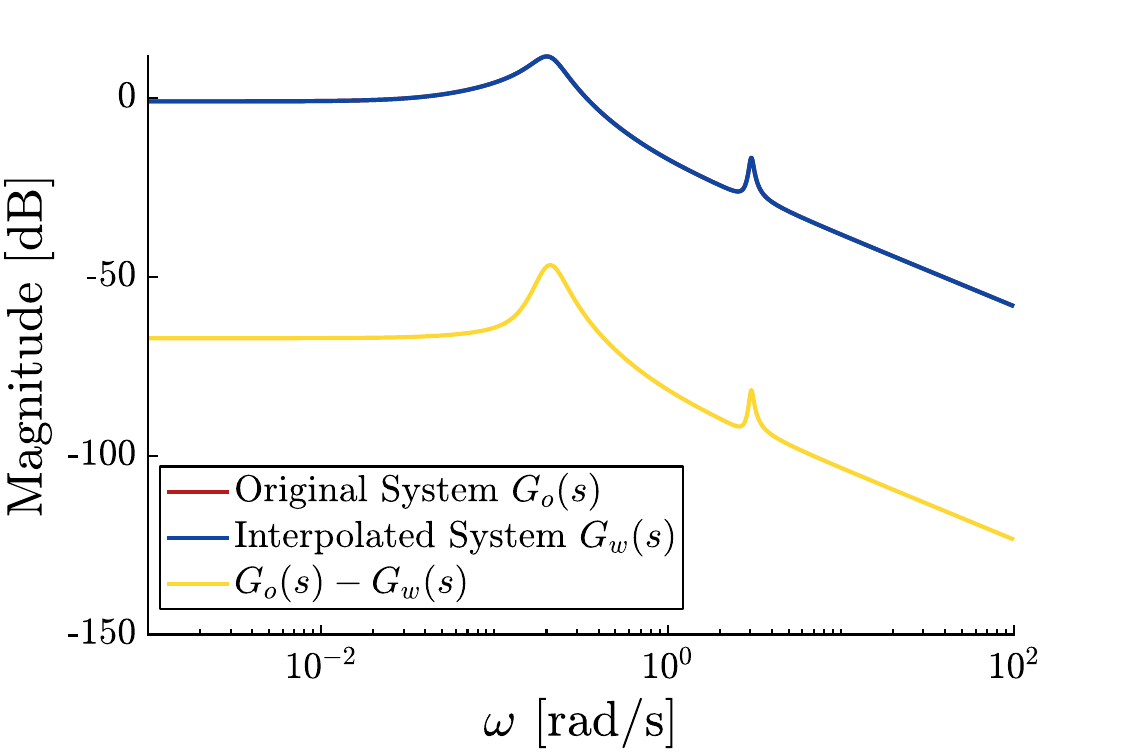}
\caption{Close prediction at $w=22$~[m/s] between blade pitch and generator speed.}
\label{fig:tf-good} 
\end{subfigure}%
\hspace{0.015\textwidth}%
\begin{subfigure}[t]{0.32\textwidth}
\centering
\noindent\includegraphics[scale=0.35]{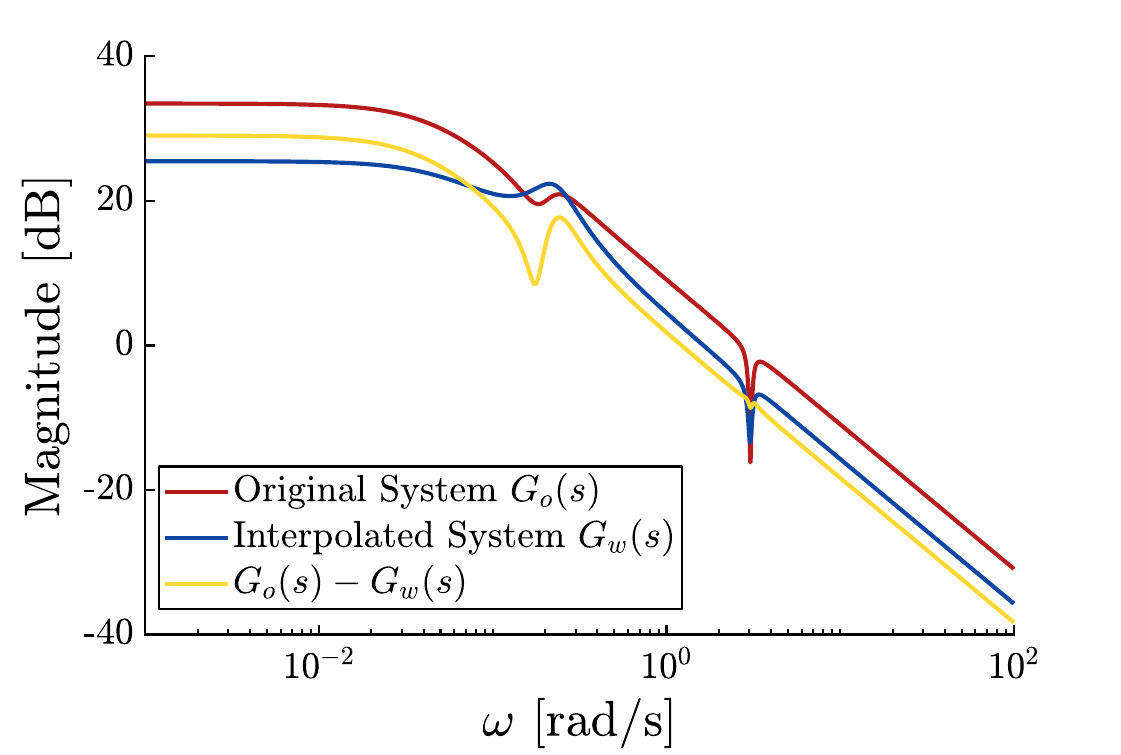}
\caption{Largest $H_\infty$ error at $w=12$~[m/s] between blade pitch and generator speed.}
\label{fig:tf-bad} 
\end{subfigure}%

\caption{Transfer function-based comparisons using the validation wind speed values for the IEA 15-MW wind turbine.}
\label{fig:interpolated-tfs} 
\end{figure*}

With a selected $\bm{W}$ (23 distinct wind speeds), the set of linearized state-space models $\Sigma_o$ at each of the wind speed values are obtained.
To construct the continuous $\Sigma_w$ using $\bm{W}$ and $\Sigma_o$, direct element-wise interpolation of the matrices $(\bm{A}_{o},\bm{B}_{o},\bm{C}_{o},\bm{D}_{o})$ was used.
To reduce the interpolation costs, matrix sparsity patterns were considered.
Only entries with nonzero values were interpolated (and the sparsity pattern remained similar \ref{test:matrixsparsity}).

To understand the predictive accuracy of this approach and check if these models satisfy \ref{test:MatrixIntp}, the following test is carried out.
Every alternate point in $\bm{W}$ was chosen as training data for the interpolation procedure, and the values in between were selected as validation points.
This allows us to assess if the interpolation approach can predict matrix properties by comparing to the validation systems\footnote{All points in $\bm{W}$ are used in the studies in Sec.~\ref{sec:results}.\vspace{0.03in}}.
In Fig.~\ref{fig:OP}, several key $\bm{\xi}_o(w)$ and $\bm{u}_o(w)$ values are shown, and there is good agreement between the interpolated LPV system and the validation points, even in the transition region.
In Fig.~\ref{fig:IntMatA}, one of the eigenvalues of $\bm{A}(w)$ that changes with the wind is shown.
Again, the eigenvalues generally are well predicted, with the largest errors in the transition region.
Finally, the normalized nonzero entries of $\bm{B}(w)$ are shown in Fig.~\ref{fig:IntMatB}.
There are some validation points with high errors in the transition region but good agreement in the other regions.

%------------------------------------
\subsection{Frequency-Domain Verification}
\label{sec: FD-val}

The transfer function matrix of the interpolated linear models was studied to understand better if the input/output relationship is accurately predicted and compute the error in Eq.~(\ref{eq:Hinf}) in \ref{test:tf}.
Here, we consider the four relationships between the two key states ($\omega_g$ and $\Theta_p$) and the inputs $\bm{u}$.
The results for the input/output combination with the highest error ($\omega_g$ and $\beta$) are shown in Fig.~\ref{fig:interpolated-tfs}.

% new paragraph
The $H_{\infty}$ norm error between the training and validation systems and the interpolated systems is shown in Fig.~\ref{fig:Hinf}.
The errors at the training points are near zero, as expected using interpolation.
However, the systems derived from the transition region~($8$--$12$~[m/s]) have the highest error compared to the other regions.
This figure shows how advanced sampling strategies could be used to better sample from regions of high error.
Additionally, the transfer functions between $\beta$ and $\omega_g$ are shown in Figs.~\ref{fig:tf-good} and \ref{fig:tf-bad} with a close prediction and largest $H_{\infty}$ error, respectively.

%------------------------------------
% \clearpage
\subsection{Time-Domain Verification}
\label{sec:time-domain-ver}

\begin{figure*}[h]
\centering
\begin{subfigure}[t]{0.32\textwidth}
\centering
\noindent\includegraphics[scale=0.35]{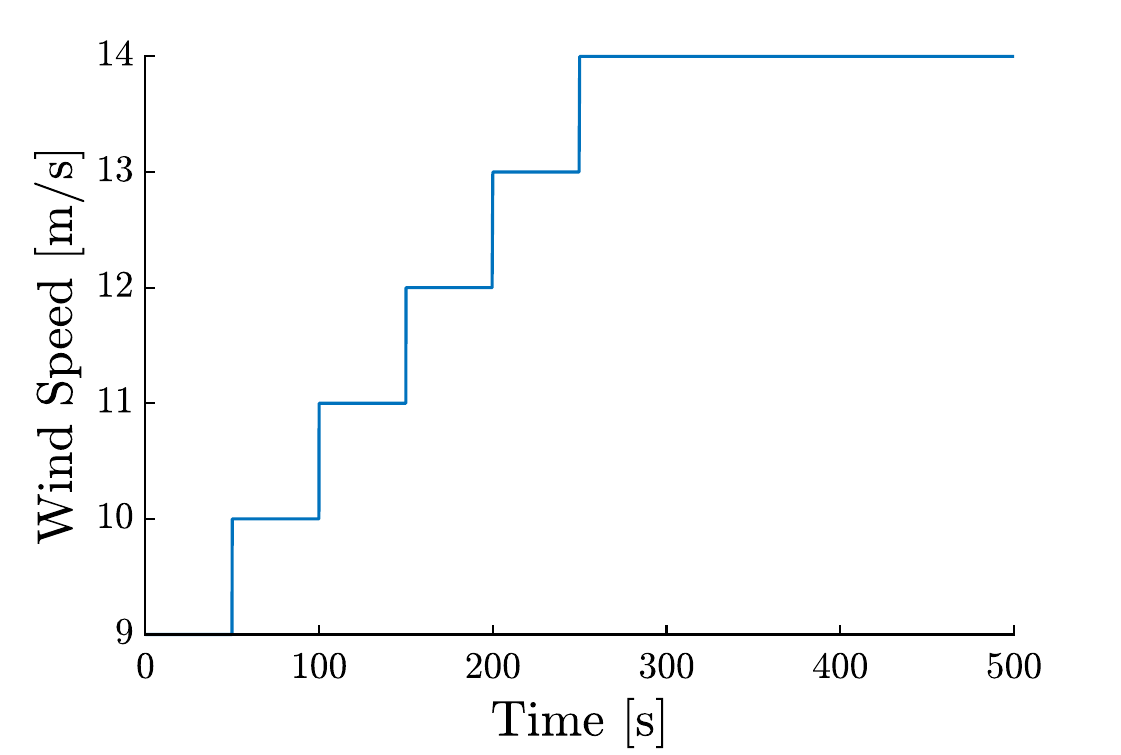}
\caption{Wind speed.}
\label{fig:resultS1WS} 
\end{subfigure}
\begin{subfigure}[t]{0.32\textwidth}
\centering
\noindent\includegraphics[scale=0.35]{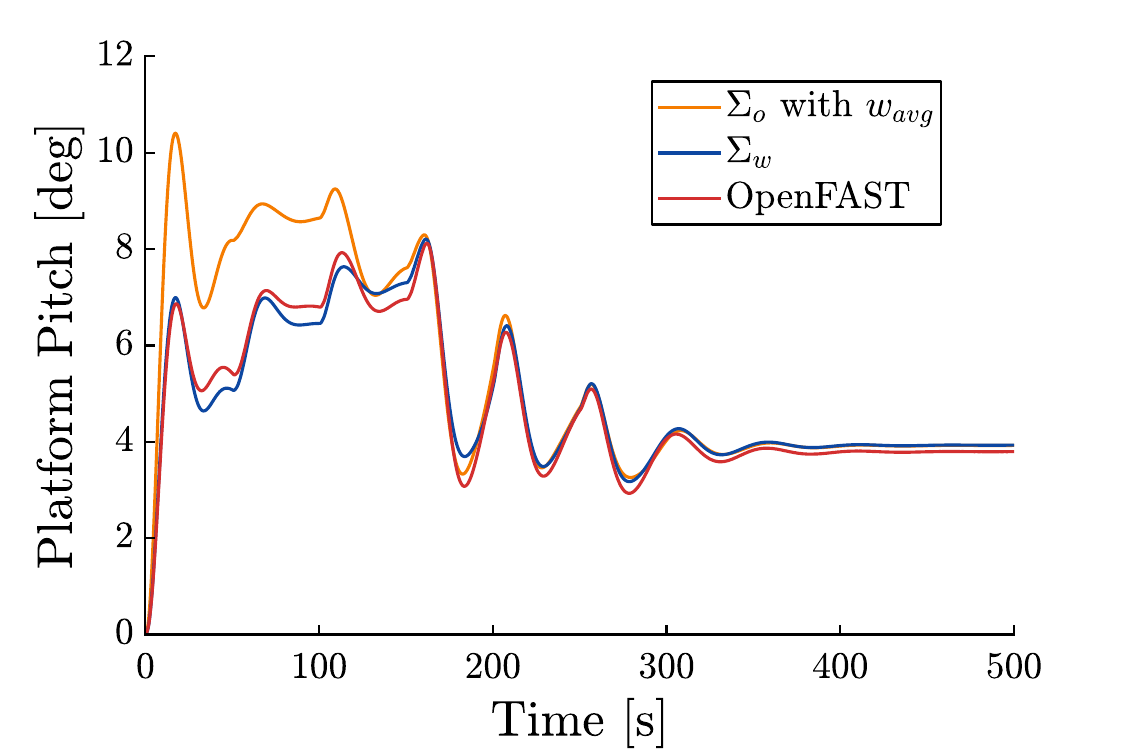}
\caption{Platform pitch.}
\label{fig:resultS1PP} 
\end{subfigure}
\begin{subfigure}[t]{0.32\textwidth}
\centering
\noindent\includegraphics[scale=0.35]{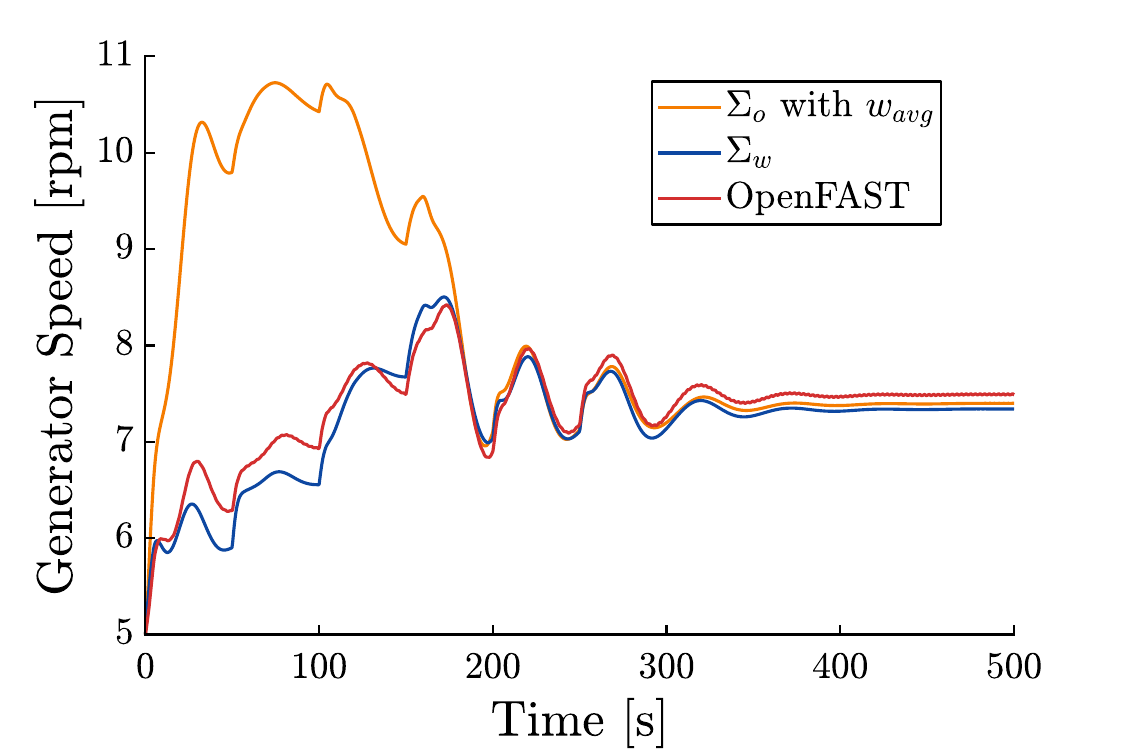}
\caption{Generator speed.}
\label{fig:resultS1GS} 
\end{subfigure} 
\caption{Model validation simulations between nonlinear $\Sigma$, LPV $\Sigma_w$, and LTI $\Sigma_o$, using $w_{\textrm{avg}}$ models.}
\end{figure*}

The final comparisons were based on \ref{test:TDSimulation} using \openFAST{} to determine the nonlinear response of $\Sigma$.
Using the same input trajectory, three different models ($\Sigma$, $\Sigma_w$, and $\Sigma_o$ using the average wind speed $w_{\textrm{avg}}$) are simulated, then the resulting state trajectories are compared.
A step-like wind input is considered for this study and is shown in Fig.~\ref{fig:resultS1WS} (and the nonzero trajectories for $\tau_g$ and $\beta$ are not shown).

From the results, we see that $\Sigma_w$ captures the nonlinear response from \openFAST{} more accurately that $\Sigma_o$ using $w_{\textrm{avg}}$.
For this study, $w_{\textrm{avg}} = 12.8$ [m/s].
Early in the simulation, when the wind speed value is significantly different from $w_{\textrm{avg}}$, we see that the $\Sigma_o$ using $w_{\textrm{avg}}$ produces inaccurate results for $\Theta_p$ in  Fig.~\ref{fig:resultS1PP} and $\omega_g$ in Fig.~\ref{fig:resultS1GS}.
Using all the different comparisons, it was concluded that the LPV model $\Sigma_w$ can, with reasonable accuracy, capture the dynamics of the considered FOWT.

\subsection{Interpolation Based on Plant Variables}
\label{sec:XpInterp}

\begin{figure*}[h]
\centering\
\includegraphics[width = 0.35\textwidth]{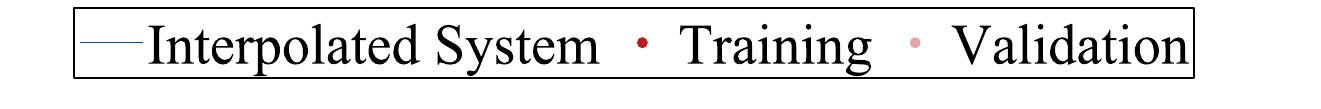}\\
\begin{subfigure}[t]{0.25\textwidth}
\centering
\noindent\includegraphics[scale=0.25]{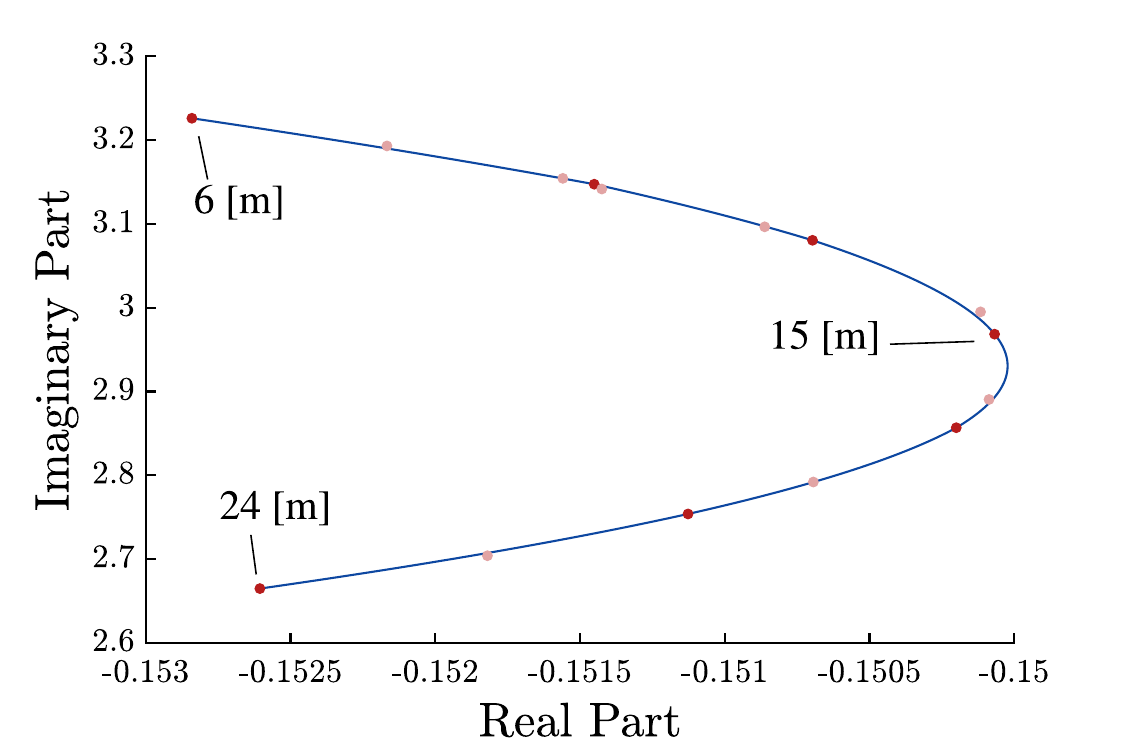}
\caption{One eigenvalue of $\bm{A}(c_d)$.}
\label{fig:eigBV} 
\end{subfigure}%
\begin{subfigure}[t]{0.25\textwidth}
\centering
\noindent\includegraphics[scale=0.25]{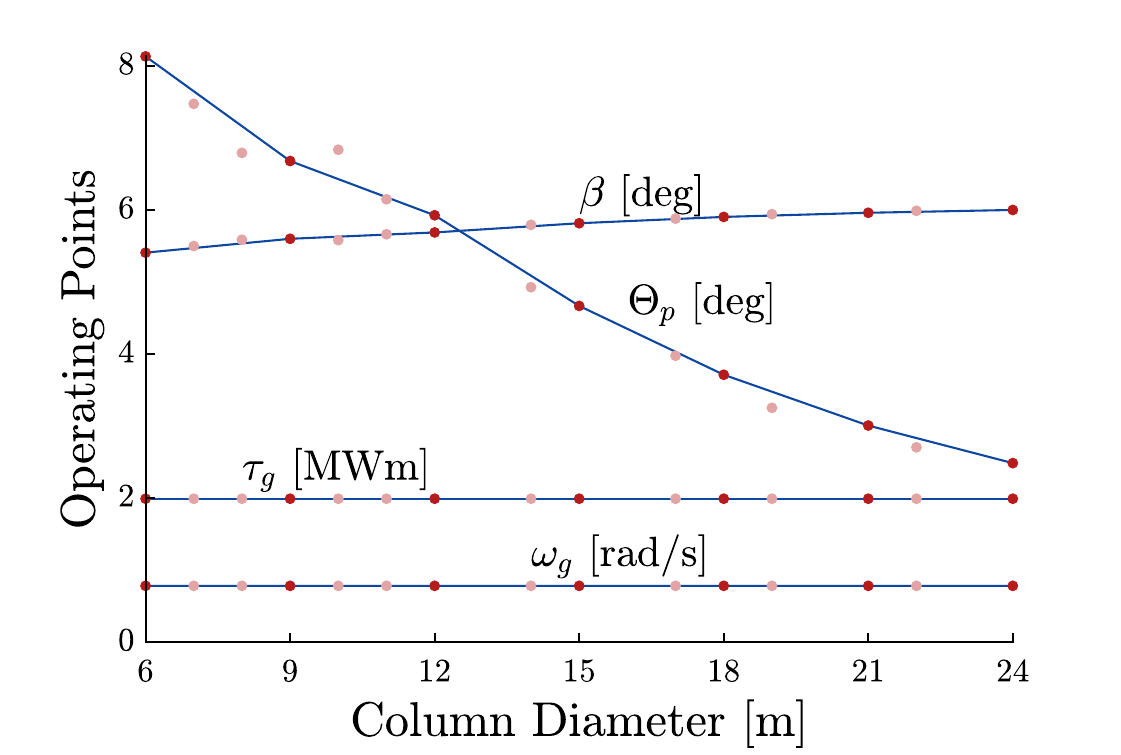}
\caption{Select stationary operating point values.}
\label{fig:OP-CD} 
\end{subfigure}%
\begin{subfigure}[t]{0.25\textwidth}
\centering
\noindent\includegraphics[scale=0.25]{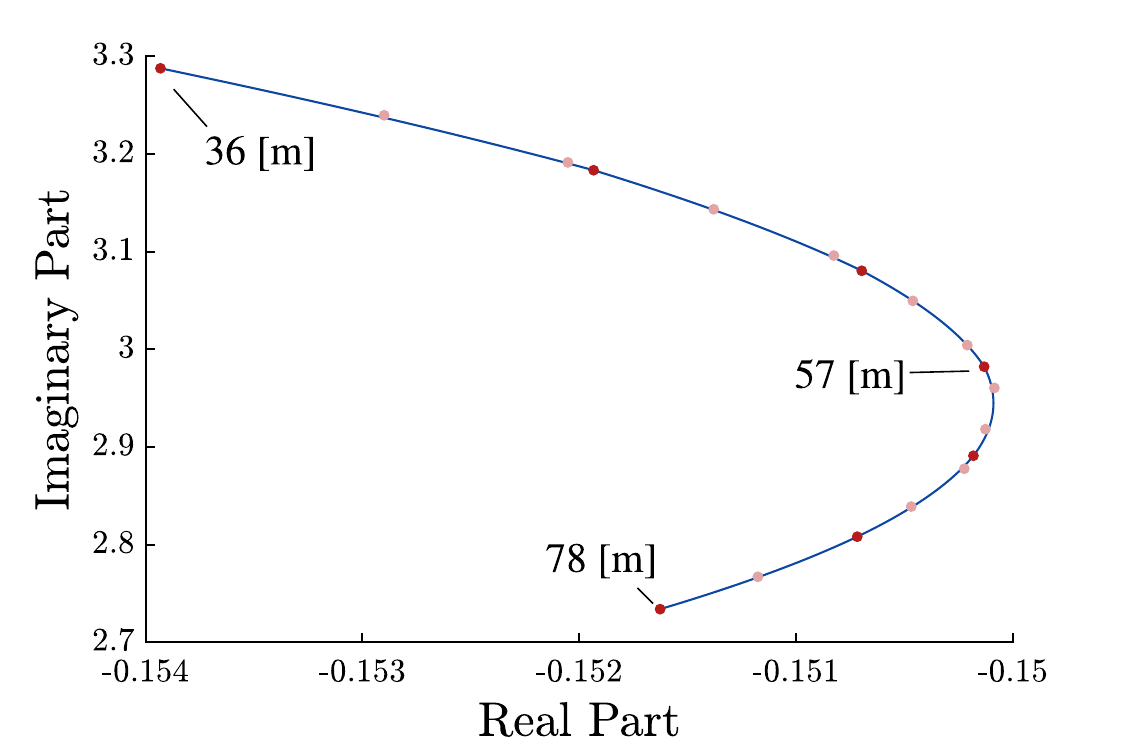}
\caption{One eigenvalue of $\bm{A}(c_s)$.}
\label{fig:eigCS} 
\end{subfigure}%
\begin{subfigure}[t]{0.25\textwidth}
\centering
\noindent\includegraphics[scale=0.25]{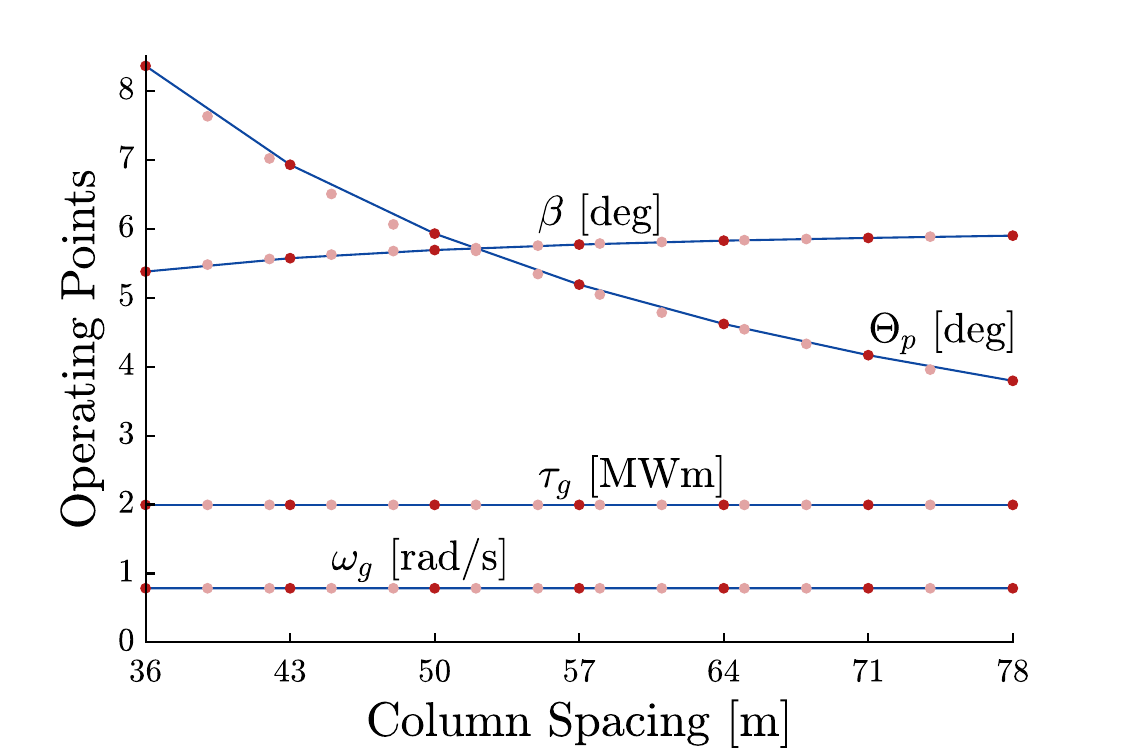}
\caption{Select stationary operating point values.}
\label{fig:OP-CS} 
\end{subfigure}%
%\noindent \includegraphics[width = \textwidth]{figures/S3/figXpVAL/plantval_combined.pdf}
\caption{Interpolation of select stationary points and eigenvalues for $\Sigma_{o}$ with $w = 12$ [m/s] based on $\bm{x}_p$.}
\label{fig:XpVal}
\end{figure*}

%\input{figures/S3/figXpVAL/plant-validation-results}

% new paragraph
The model $\Sigma_w$ just presented was obtained using a particular instance of the system's plant design, denoted by $\bm{x}_p$ in Eq.~(\ref{eq:plant_var}).
However, we also want to consider the design impacts of the plant variables over the full range of their allowable values.
For such an investigation, a complete set of linear models $\bm{\Sigma}_w$, corresponding to multiple plant designs, are obtained.
Because only two plant variables are considered in the study, a full-factorial grid was constructed. 
A regular-grid interpolation scheme is then used to interpolate the individual elements of $\Sigma_w$ over the entire range of the plant variables.
From \ref{test:matrixsparsity} the sparsity information can be used to construct the interpolation scheme for the nonzero elements in the linear system matrices, making the process more efficient.
The samples were generated between bounds $\bm{L}_p = [36,6]^T$ [m] and $\bm{U}_p = [78,24]^T$ [m] considered for $c_s$ and $c_d$ dimensions, respectively.
The column spacing dimension was sampled for $n_{cs} = 7$ different values, while the column diameter was sampled at $n_{cd} = 7$ different values, yielding a total of $n = 49$ samples.
The nominal platform specifications are available at~\cite{Allen2020}.

Similar tests to those outlined in Secs.~\ref{sec:SS-model-val}-\ref{sec: FD-val} were carried out to check the predictive accuracy of the interpolation scheme based on $\bm{x}_p$. 
For the state-space model comparison, the interpolation scheme was set up for the linear models with the highest $H_{\infty}$ error from Fig.~\ref{fig:tf-bad} at $w = 12$ [m/s].
The corresponding state matrix $\bm{A}(12)$ and key states and control operating points from $\{\bm{\xi}_o(12), \bm{u}_o(12)\}$ were interpolated individually for both column spacing ($c_s$) and column diameter ($c_d$) dimensions as shown in Fig.~\ref{fig:XpVal}.
The frequency domain validation outlined for interpolation based on $\bm{W}$ was carried out for interpolation based on $\bm{x}_p$.
The $H_{\infty}$ norm error for $25$ different plant variable samples was evaluated between the interpolated and actual models at $w = 12$ [m/s], and the average error was found to be $\sim 10^{-5}$ [dB] for all 25 samples.
Therefore, we conclude that interpolation based on $\bm{x}_p$ is generally well-behaved, potentially more so than the wind speed dimension.
Since the $H_{\infty}$ error was so low, these results are not shown graphically.
%---------------------------------------------------------------
% Formulation
%------------------------------------------
\xsection{Control Co-Design Problem formulation }
\label{sec:OCP}

This section describes the nested CCD problem constructed using the LPV models from Sec.~\ref{sec:LPV} to study the impact of various stability constraints on the LCOE for the considered single-device FOWT.

%-------------------------------------
\subsection{Outer-Loop Plant Design Problem Formulation}
\label{sec:PlantDesign}

The outer-loop plant optimization problem in the nested CCD approach employed here is centered around the LCOE calculation in Eq.~(\ref{eq:LCOE}).
In this calculation, the total lifetime cost is estimated as:
\begin{subequations}
\begin{align}
\label{eq:cost}
C_{\textrm{capital}}(\bm{x}_p) &= C_{\textrm{turbine}}(\bm{x}_p) + C_{\textrm{bos}}(\bm{x}_p) \\
C_n &= r_{\textrm{fc}}C_{\textrm{capital}}(\bm{x}_p) + C_{\textrm{opex}} 
\end{align}
\end{subequations}

\noindent where $C_{\textrm{turbine}}(\bm{x}_p)$ and $C_{\textrm{bos}}(\bm{x}_p)$ are the turbine cost and the balance of system cost for the turbine that depends on the plant design.
$C_{\textrm{opex}}$ is the annual operating costs, and $r_{\textrm{fc}}$ is the fixed charge rate, which, as used in this study, captures the amortization of $C_{\textrm{capital}}$ in Eq.~(\ref{eq:cost}) across the project lifetime.
More details about $r_{\textrm{fc}}$ can be found in~\cite{Theis2021, Short1995}. 
For this study, we used the cost and scaling models and LCOE equation discussed in detail in~\cite{WISDEM, Fingersh2006, Malcolm2006, Stehly2021, Mowers2021,GarciaSanz2019b}.

The total energy generated in a year is determined as:
\begin{align}
\label{eq:Eweibull}
E =  \text{AEP} = \int_{\mathcal{W}_{o}} \bar{P}^{*}(\bm{w}(t,W_o),\bm{x}_p)\bm{f}_{\tilde{W}_{o}}(W_o)\, \mathrm{d}W_o
\end{align}
\noindent where $\mathcal{W}_o$ is the entire operating region, $\bm{w}(\cdot)$ is a given load case with an average wind speed of $W_o$, $\bar{P}^{*}(\cdot)$ is the average power produced for a given plant design and design load case (DLC), and $\bm{f}_{\tilde{W}_{o}}$ is the Weibull probability density function that describes the wind speed distribution.
Eleven wind profiles from the IEA-specified DLCs with the normal turbulence model in~\cite{DNV2016} (i.e.,~`DLC 1.1') with average wind speed values between $3$ and
$25$ [m/s] are used to approximate the distribution $\bm{f}_{\tilde{W}_{o}}$.

Finally, the annual energy production (AEP) is calculated as:
\begin{align}
   E_n =  (1-f_{wl}) E
\end{align}

\noindent where $0\leq f_{wl} \leq 1$ is the wake loss factor.
Both $C_n$ and $E_n$ are normalized with respect to the machine rating, which is $15$ [MW].
This operation does not change the value of the LCOE, but it changes the units of $C_n$ to [\$/\textrm{MW}] and $E_n$ to [h].
Therefore, LCOE = $C_n$/$E_n$, and the complete outer-loop optimization problem is:
\begin{subequations}
\begin{align}
\min_{\bm{x}_p} \quad & \text{LCOE}(\bm{x}_p) \\
\text{subject to:}\quad & \bm{L}_p \leq \bm{x}_p \leq \bm{U}_p
\end{align}
\end{subequations}

\noindent where only simple upper and lower bounds on the plant variables are considered at this time (although more complex plant-only constraints can be readily incorporated).
Note that for a fixed plant $\bm{x}_p^{\dagger}$, the solution for each $\bar{P}^{*}(\bm{x}_p^{\dagger},\bm{w})$ can be determined through independent minimization problems.
Therefore, the control subproblems can be solved in parallel.

%------------------------------------------------
\subsection{Control Subproblem for a Specific Design Load Case}
\label{sec:ControlSubproblem}

%\subsection{Control Subproblem}
The control subproblem's goal is to understand the impact of the control decisions on system response, power production, and ultimately the LCOE design objective.
An open-loop optimal control problem is constructed to maximize the power produced for a given operational scenario or DLC. 
The optimization formulation is presented using the original notation for states and controls $(\bm{\xi},\bm{u})$, but the linear time-varying transformation in Eq.~(\ref{eq:operatingpoint}) based on the wind-dependent operating point is applied so that $(\bm{\xi}_{\Delta},\bm{u}_{\Delta})$ are the states and controls for this subproblem.

The energy produced by the turbine is:

\begin{align}
\label{eq:power}
\int_{0}^{t_f} P(t) \mathrm{d}t = \int_{0}^{t_f}  \eta_g \tau_g(t) \omega_g(t) \mathrm{d}t
\end{align}

\noindent where $\eta_g$ is the generator efficiency.
Note, the control term $\tau_g$ appears linearly in the objective term Eq.~(\ref{eq:power}).
The presence of linear control terms in the objective function with linear dynamics can give rise to singular arcs~\cite{Betts2010} as the control trajectory cannot be uniquely determined.
To help mitigate this issue, a quadratic penalty term is introduced in the objective term:
\begin{align}
\label{eq:powerCor}
\Pi_c(t) =  \bm{u}^T\begin{bmatrix}
10^{-16} & 0\\
0 & 10 \end{bmatrix}
\bm{u}
\end{align}

\noindent where values in this penalty matrix were identified according to the method discussed in~\cite{Herber2014a}.
In addition to this, a penalty is added to limit the fluctuation of the platform pitch:
\begin{align}
\label{eq:pitchPen}
\Pi_p(t) = 
\Theta_p^2
\end{align}

\begin{table}[t]
\renewcommand{\arraystretch}{1}
\begin{center}
\caption{CCD study parameters.}
\label{tab:probparameters}
\begin{tabular}{rll}
  \hline\hline
 {Variable} & {Value} & {Units} \\ 
  \hline
$\omega_{g,\max,1}$ & 0.7850 & [rad/s] \\ 
$\omega_{g,\max,2}$ & 0.9424 & [rad/s] \\ 
$\Theta_{p,\max}$ & 6 & [deg] \\ 
%$P_{\max}$ & 15 & [MW] \\ 
$\tau_{g,\max}$& 19.8 & [MNm]\\
$F_{s,\max}$& 5000& [kN]\\
$M_{s,\max}$&32000&[kNm]\\
$\beta_{\max}$&0.3948&[rad]\\
$r_{\textrm{fc}}$&0.056&-\\
$f_{\textrm{wl}}$&0.15&-\\
\hline \hline
\end{tabular}
\end{center}
\end{table}

The linear dynamic constraints included using $\Sigma_w$ from Eq.~(\ref{eq:sigma-w}) with plant dependence are:%
\begin{align}
\label{eq:dynamics}
\frac{d\bm{\xi}_{\Delta}}{dt} = 
\bm{A}(\bm{x}_p,w) \bm{\xi}_{\Delta}
+ \bm{B}(\bm{x}_p,w) \bm{u}_{\Delta}
- \displaystyle\frac{\partial \bm{\xi}_o(\bm{x}_p,w)}{\partial w} \frac{d w}{d t}
\end{align}

\noindent and the initial state values correspond to the state operating points for $w(0)$:
\begin{align}
\label{eq:InitialStates}
\bm{\xi}(0) = \bm{\xi}_o(w(0)), \text{ or equivalently }  \bm{\xi}_{\Delta}(0) = \bm{0}
\end{align}

To protect the generator components from excess electrical loads and the nacelle from the dynamic loads, an upper bound for generator speed $\omega_g$ is set restricting the speed to the rated speed of the turbine:
\begin{align}
\label{eq:UBGenSpeed}
0 \leq \omega_g(t) \leq \omega_{g,\max}
\end{align}

\noindent As a proxy for the stability and safety of the FOWT system, an upper bound on the platform pitch tilt $\Theta_p$ is included:
\begin{align}
\label{eq:UBPtfmPitch}
\Theta_p(t) \leq \Theta_{p,\max} 
\end{align}

\noindent Maximum and minimum value constraints are placed on the controls blade pitch $\beta$ and the generator torque $\tau_g$, according to the values prescribed in~\cite{Gaertner2020}:
\begin{subequations}
\label{eq:Controlconstraints}
\begin{gather}
0 \leq \tau_g(t) \leq \tau_{g,\max} \\
0 \leq \beta(t) \leq \beta_{\max}
\end{gather} % DRH: made 0 the limits
\end{subequations}

Using the model for outputs from Eq.~(\ref{eq:sigma-w}), we include additional output constraints on tower base fore-aft shear force and tower base side-to-side moment, respectively:
\begin{subequations}
\label{eq:outputconstraints}
\begin{gather}
F_{s} \leq F_{s,\max} \\
M_{s} \leq M_{s,\max}
\end{gather} % DRH: made 0 the limits
\end{subequations}

The complete control subproblem formulation is presented in Problem~(\ref{eq:OC}), and solved with weight $k = 10^{-8}$ to normalize the objective function value to be approximately unity magnitude:
\begin{subequations}
\label{eq:OC}
\begin{align}
\min_{\bm{u}_{\Delta},\bm{\xi}_{\Delta}} \quad & \int_{0}^{t_f} \left( -kP(t) + \Pi_c(t) + \Pi_p(t) \right) \mathrm{d}t \\
\text{subject to:}\quad &  \text{Eqs.~}(\ref{eq:dynamics}) - (\ref{eq:outputconstraints})
\end{align}
\end{subequations}

\noindent which will yield the average power $\bar{P}^* = \left(\int_{0}^{t_f} P(t) \mathrm{d}t\right)/t_f$ needed in Eq.~(\ref{eq:Eweibull}).
It can be observed that Problem~(\ref{eq:OC}) has only quadratic objective function terms and linear constraints; therefore, it can be classified as a LQDO problem (see Sec.~\ref{sec:Nested}).
%----------------------------------------------------------------
% results
\xsection{Results}\label{sec:results}

\begin{figure*}[t]
\centering
\includegraphics[width = 0.7\textwidth]{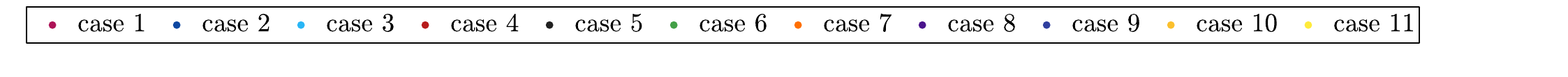}\\
\begin{subfigure}[t]{0.32\textwidth}
\centering
\noindent\includegraphics[scale=0.35]{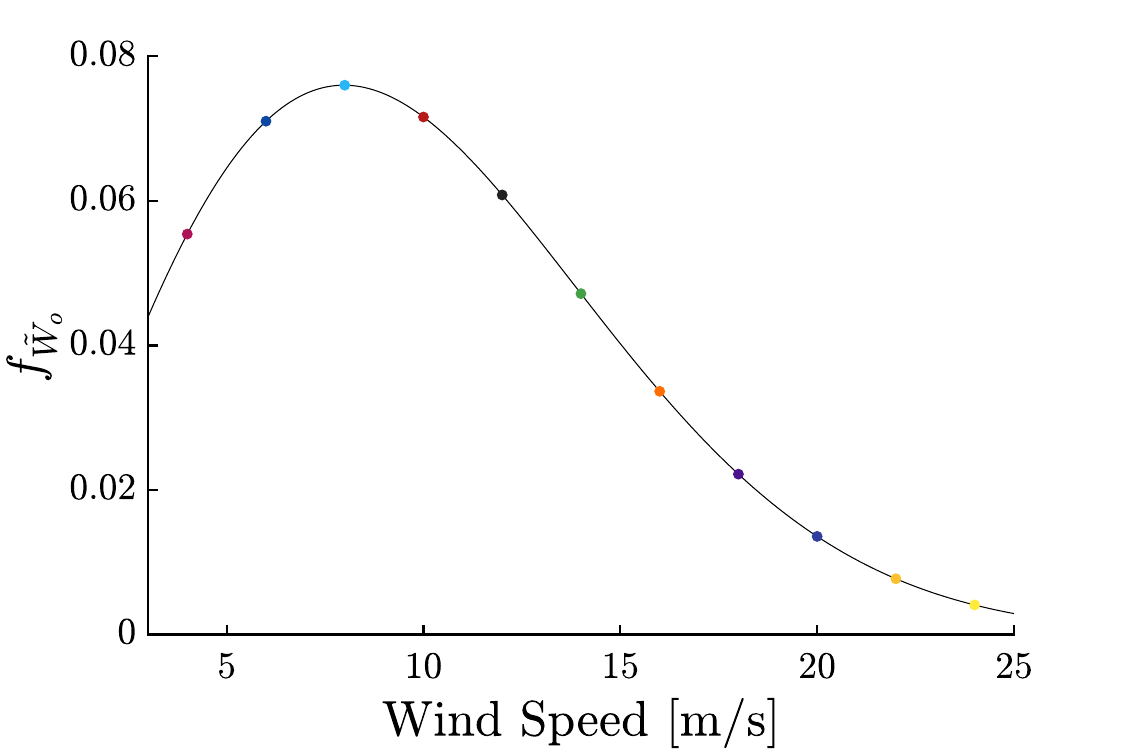}
\caption{Weibull PDF with $k=2$ and $\lambda = 11.28$.}
\label{fig:DLC_weight} 
\end{subfigure}%
%----------------------------
\begin{subfigure}[t]{0.32\textwidth}
\centering
\noindent\includegraphics[scale=0.35]{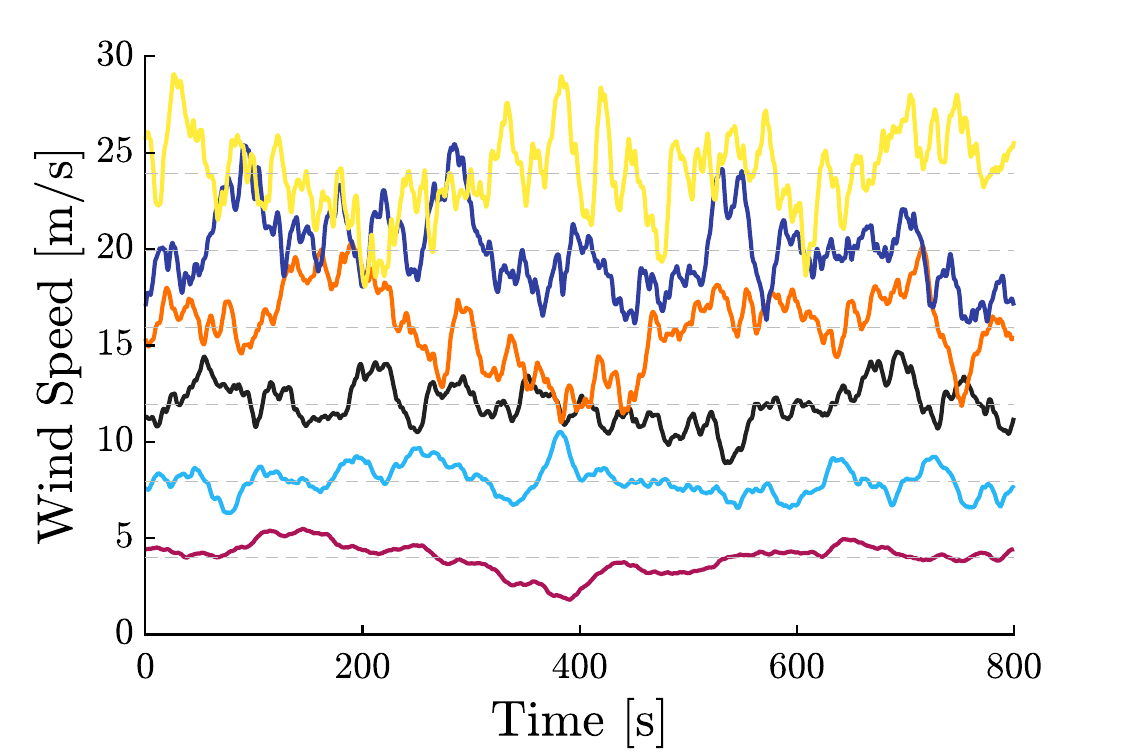}\textbf{}
\caption{Cases \{1,3,5,7,9,11\}.}
\label{fig:DLCr1}
\end{subfigure}%
%-------------------------
\begin{subfigure}[t]{0.32\textwidth}
\centering
\noindent\includegraphics[scale=0.35]{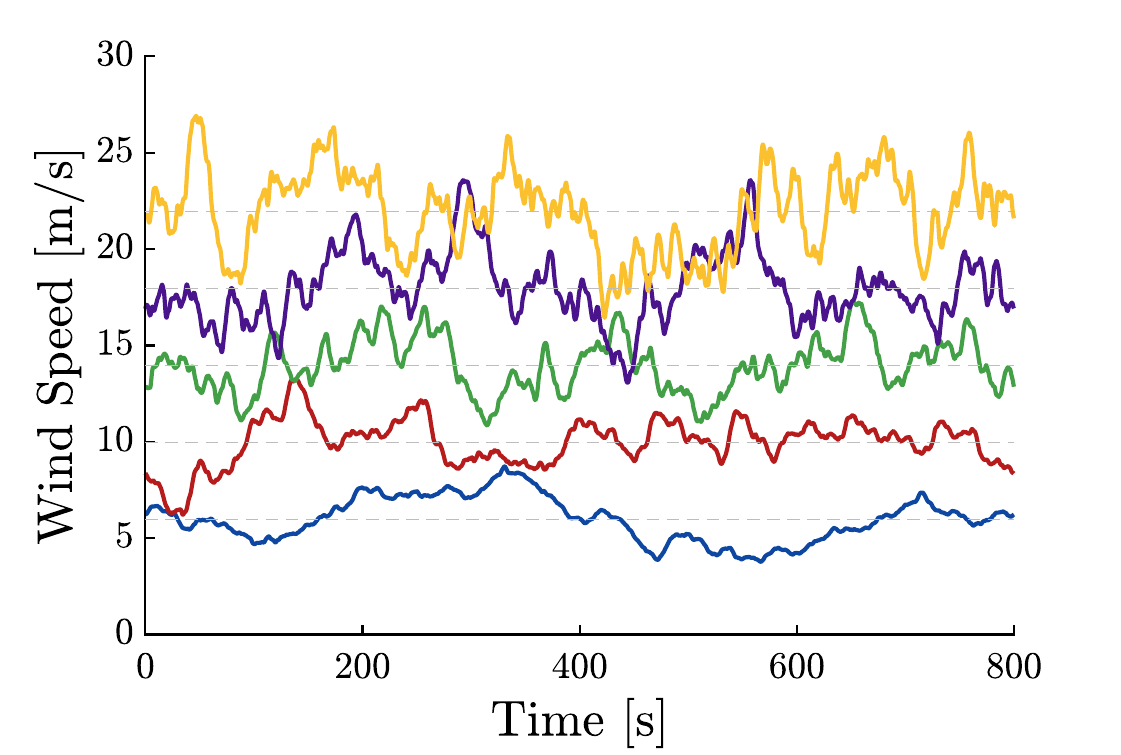}
\caption{Cases \{2,4,6,8,10\}.}
\label{fig:DLCr2} 
\end{subfigure}%
%-------------------------
%\noindent \includegraphics[width = \textwidth]{figures/S5/figDLC/DLC_combined.pdf}
\caption{Input wind profiles from DLC 1.1. based on the average wind speed for the trajectory.}
\label{fig:windDLC}
\end{figure*}

In this section, we describe the results of an LCOE-focused CCD study using the IEA 15-MW wind turbine~\cite{Gaertner2020} supported by a floating semisubmersible platform~\cite{Allen2020}.
The values for the CCD problem parameters defined in Sec.~\ref{sec:ControlSubproblem} are given in Table~\ref{tab:probparameters}.
Here, we consider the following 11 wind load cases based on the average input wind speed shown in Fig.~\ref{fig:windDLC}.
Extrapolation is used to find the values of the LPV model $\Sigma_w$ outside the $3$--$25$ m/s range, because the models are readily predictable in these regions.

The LQDO problems of the form in Problem~(\ref{eq:OC}) are solved using \texttt{DTQP}, an open-source \texttt{MATLAB}-based toolbox using the DT method and quadratic programming~\cite{dtqp, Herber2017e}.
Each problem was discretized using 2,500 equidistant mesh points, with an observed relative objective function error bound of approximately $10^{-4}$.
% \xchange{The LQDO problem was not scaled.}

% new paragraph (plant decisions)
A sensitivity approach was used to explore how the plant design decisions impact the system's cost and performance.
Although a hybrid-optimization scheme could be used to identify the single optimal design as shown in~\cite{SundarrajanX1}, a sensitivity study was utilized to better understand the different trade-offs.
To understand the impact of plant variables on the system stability, power production, and, subsequently, the LCOE, several constraint bounds on the platform pitch tilt $\Theta_p$ were explored.
More specifically, an exhaustive sensitivity study was conducted where $\Theta_p$ was constrained to five different values between $3^\circ$ and $7^\circ$.
A $60 \times 60$ grid was used to sample the plant design space.
Although no wave/current forces are included as disturbances at this time, these different constraint values on $\Theta_p$ will roughly indicate performance in more dynamic wave and current conditions.

\begin{figure*}
\centering
\begin{subfigure}[b]{0.32\textwidth}
    \centering
    \includegraphics[scale=0.41]{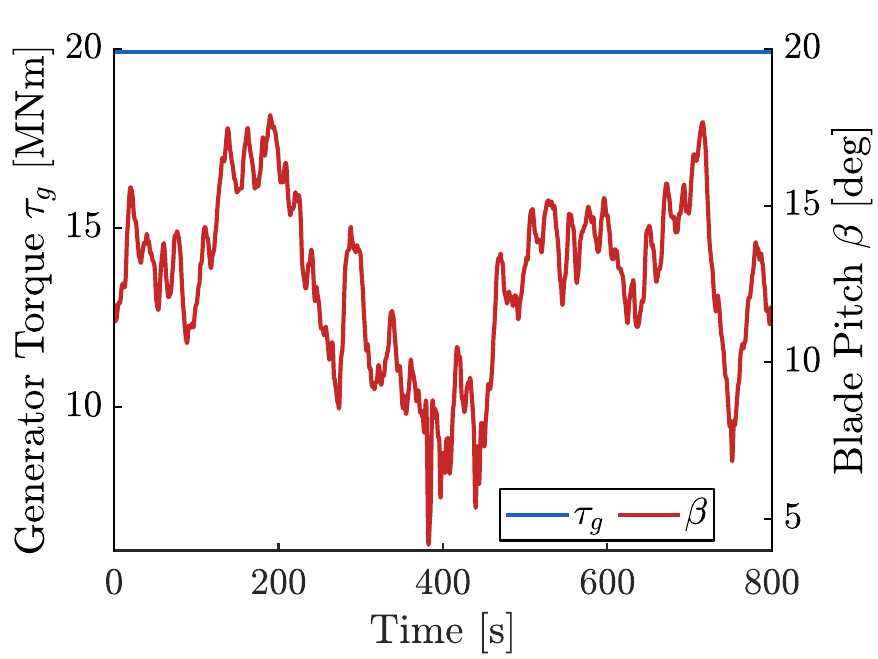}
    \caption{Select states $\omega_g$ and $\Theta_p$.}
    \label{fig:OCstates}
\end{subfigure}%
\hspace{0.01\textwidth}%
\begin{subfigure}[b]{0.32\textwidth}
    \centering
    \includegraphics[scale=0.41]{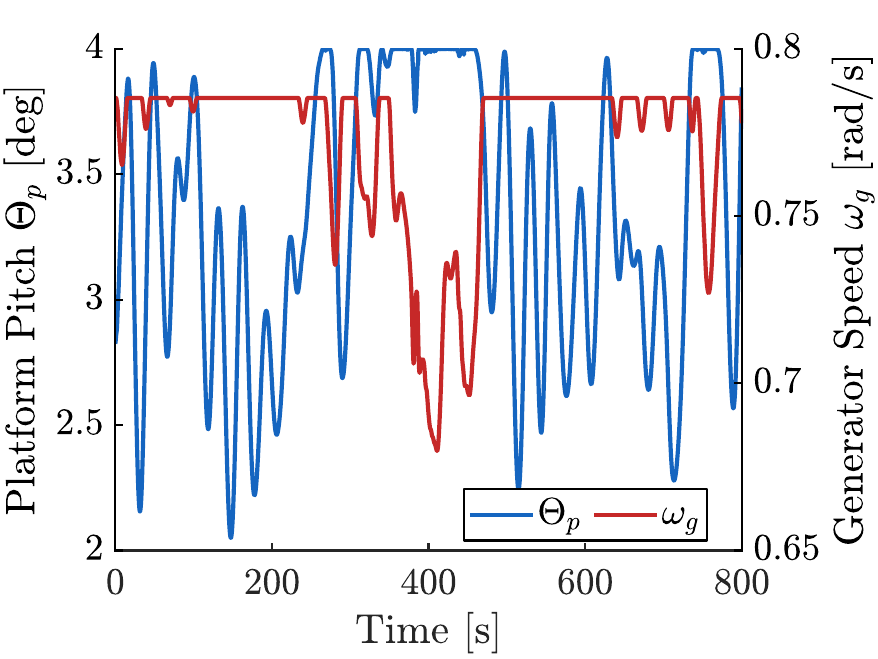}
    \caption{Controls $\beta$ and $\tau_g$.}
    \label{fig:OCcontrol}
\end{subfigure}%
\hspace{0.01\textwidth}%
\begin{subfigure}[b]{0.32\textwidth}
    \centering
    \includegraphics[scale=0.41]{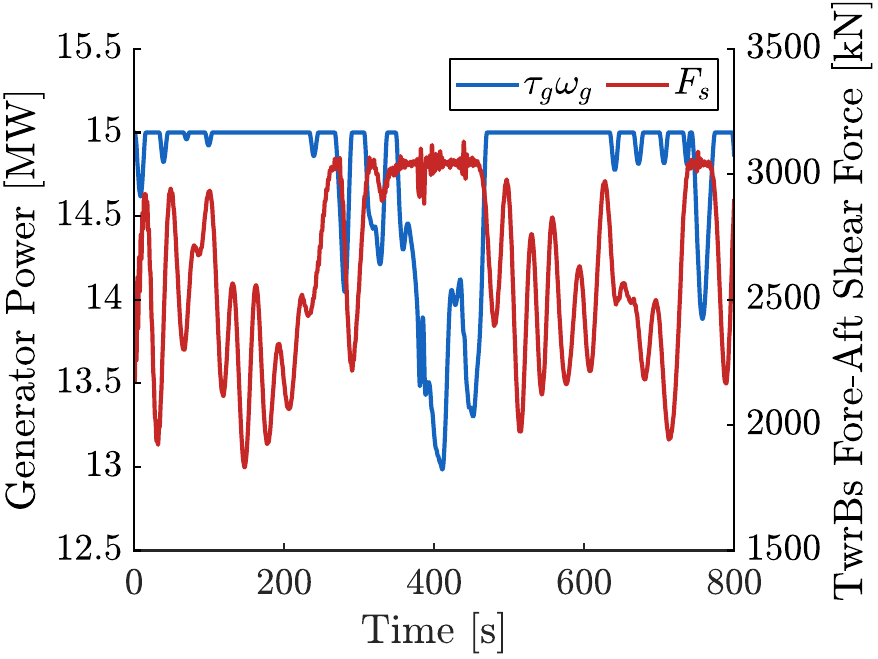}
    \caption{Select outputs Power $\tau_g\omega_g$ and $F_{s}$.}
    \label{fig:OCpower}
\end{subfigure}%
\caption{Optimal control results with nominal plant dimension $\bm{x}_{p,\textrm{nominal}}$, case 7, $\omega_{g,\max,1}$ ,and $\Theta_p\leq 4^\circ$.}
\label{fig:ocresults}
\end{figure*}

\begin{figure*}[h]
\centering
\includegraphics[width=0.4\textwidth]{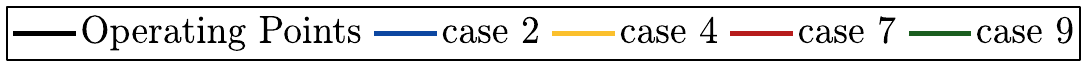}\\
\begin{subfigure}[t]{0.25\textwidth}
\centering
\noindent\includegraphics[scale=0.25]{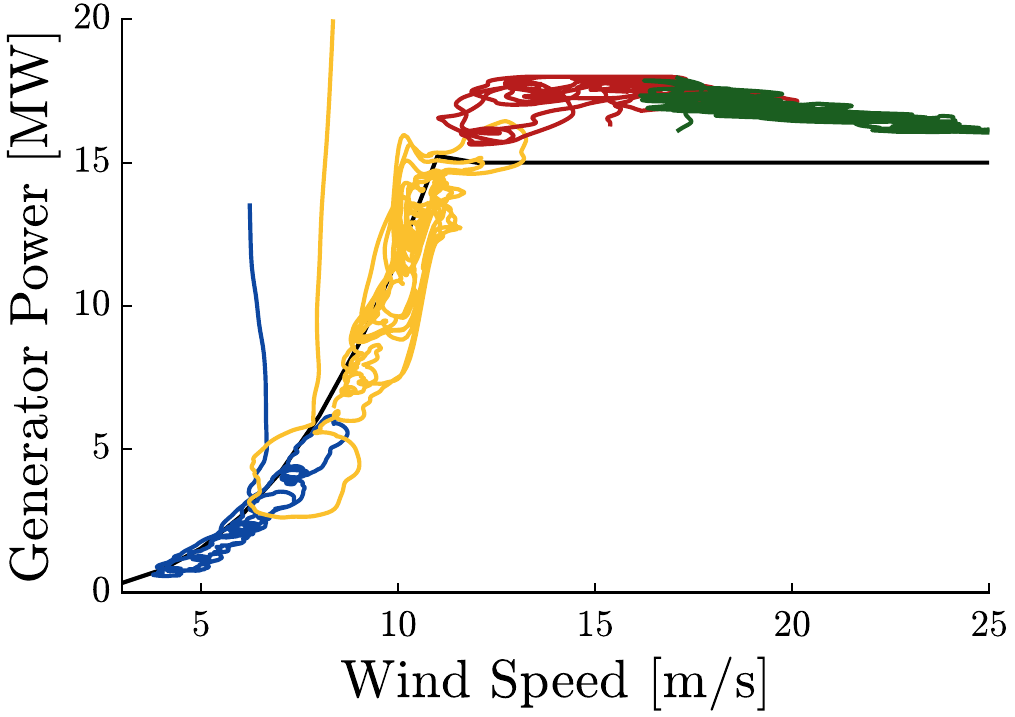}
\caption{Wind speed vs.~generator power.}
\label{fig:WSvsGP} 
\end{subfigure}%
\begin{subfigure}[t]{0.25\textwidth}
\centering
\noindent\includegraphics[scale=0.25]{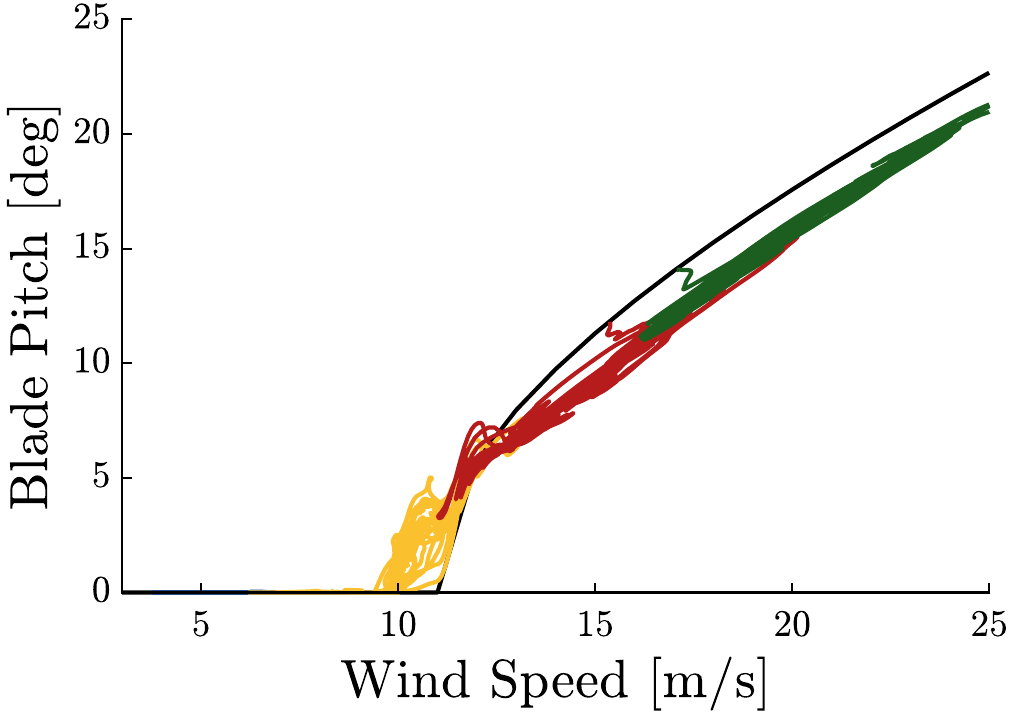}
\caption{Wind speed vs.~$\beta$.}
\label{fig:WSvsBP} 
\end{subfigure}%
\begin{subfigure}[t]{0.25\textwidth}
\centering
\noindent\includegraphics[scale=0.25]{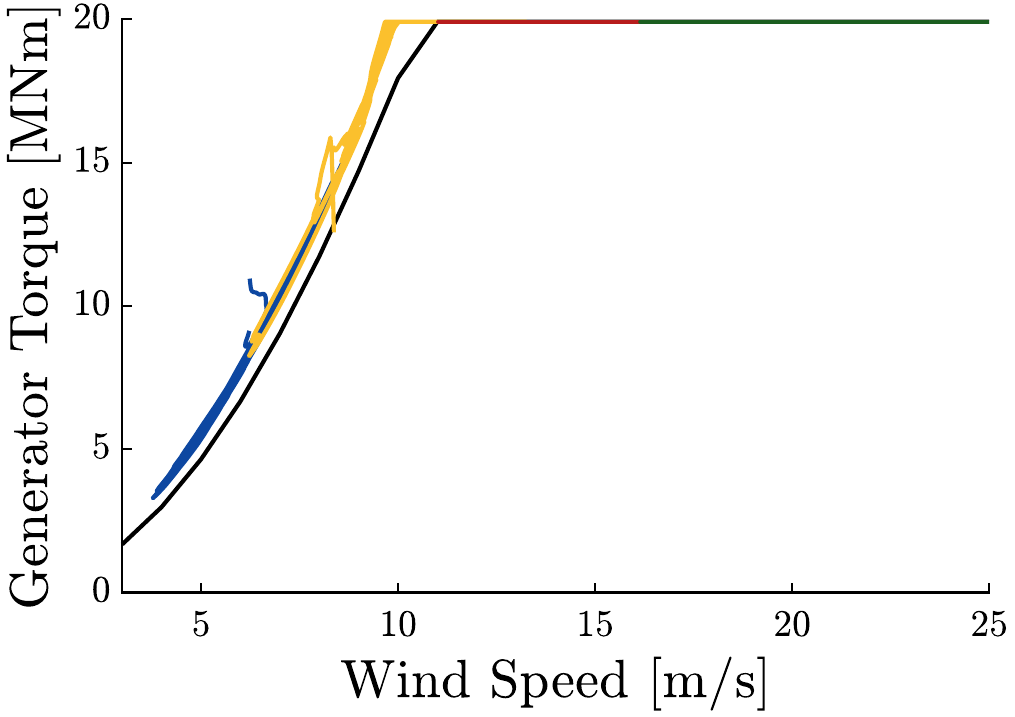}
\caption{Wind speed vs. $\tau_g$.}
\label{fig:WSvsGT}
\end{subfigure}%
\begin{subfigure}[t]{0.25\textwidth}
\centering
\noindent\includegraphics[scale=0.25]{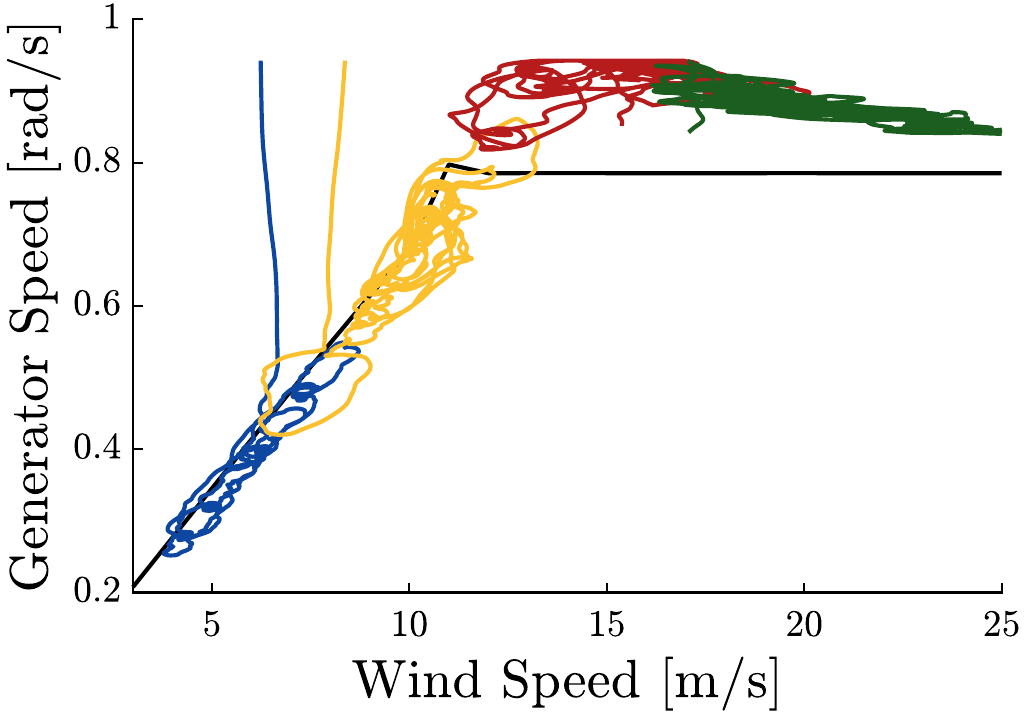}
\caption{Wind speed vs. $\omega_s$.}
\label{fig:WSvsGS}
\end{subfigure}%
\caption{Select optimal control results using the LPV model vs.~operating point schedule with $\bm{x}_{p,\textrm{nominal}}$, $\omega_{g,\max,2}$, and $\Theta_p\leq 6^\circ$. }
\label{fig:DLCvsREG}
\end{figure*}

\subsection{Notes on the Computational Time}\label{sec:computational-time-notes}
A desktop workstation with an AMD 3970X CPU, 128-GB DDR4 2,666-MHz RAM, \texttt{Matlab} 2021b update 2, and \texttt{Windows} 10 build 17763.1790 was used to obtain all the linear models and perform the different CCD studies.
The linear models were obtained using the WEIS toolkit available at~\cite{WEIS}.
Approximately 90 hours are required to obtain the complete set of linear models discussed in Sec.~\ref{sec:XpInterp}, the most computationally expensive operation in this study.
Once the linear models corresponding to the full-factorial scheme are available, 0.8 seconds are required to construct and evaluate the surrogate model.
The average solution time for constructing and solving a single inner-loop subproblem shown in Eq.~(\ref{eq:OC}) is 0.74 seconds, which includes determining physically accurate trajectories with respect to the linear model.
The average time for solving the different subproblems for all 11 load cases shown in Fig.~\ref{fig:windDLC} in parallel is 8.2 seconds. 
The computational cost to obtain the results for a single value of $\Theta_{p,\max}$ was, on average, $8.2$ hours.
Overall, there were $3,600 \times 11 \times 5 = 198,000$ inner-loop control subproblems solved for different values of plant variables $\bm{x}_p$, wind case, and $\Theta_{p,\max}$.

All the studies discussed in this paper are formulated and solved using {\sc Matlab}.
However, the code to run the inner-loop studies using the LPV models is also available in \texttt{Python} and is published as part of the WEIS tool~\cite{WEIS}.
The code for inner-loop studies mentioned in the previous sections is available in {\sc Matlab} at~\cite{dtqp}.

%----------------------------------------------
\subsection{Results for a Single-Control Subproblem}\label{subsec:single_OC}

Figure~\ref{fig:ocresults} summarizes the optimal control results for one of the 198,000 problems with nominal plant dimensions ($\bm{x}_{p,\textrm{nominal}} = [51.75,12.50]$), load case 7, maximum generator speed value of $\omega_{g,\max,1} = 0.7850$~[rad/s], and $\Theta_p\leq 4^\circ$.
The optimal trajectories for the generator speed and platform pitch are shown in Fig.~\ref{fig:OCstates}.
We see that the constraint $\Theta_p\leq 4^\circ$ and others in Table~\ref{tab:probparameters} are satisfied.
Load case 7 is in the rated region, so we might expect the blade pitch to be the primary mode of control and the generator torque to be held roughly constant~\cite{Pao2009}.
As shown in Fig.~\ref{fig:OCcontrol}, these trends are reflected in the optimal control results.
In addition to these, from Fig.~\ref{fig:OCpower}, we see that the constraints placed on the tower base fore-aft shear force in Eq.~(\ref{eq:outputconstraints}) are satisfied.
The constraint placed on the tower base side-to-side moment is also satisfied, but it is not shown.
To satisfy the platform pitch constraints, we see that the generator speed does need to decrease when the pitch constraint becomes active.
Consequently, from Figs.~\ref{fig:OCstates} and \ref{fig:OCpower}, we can see how the generator power is affected by the pitch constraint because it is a function of the generator speed.

To better understand the optimal control results in other operating regions, Fig.~\ref{fig:DLCvsREG} was constructed to show the behavior of a system with nominal plant values $\bm{x}_{p,\textrm{nominal}}$ and the pitch constraint $\Theta_p\leq 6^\circ$.
The constraint $\omega_{g,\max,1}$ was relaxed by $20\%$ to be $\omega_{g,\max,2} = 0.9424$~[rad/s] to explore solutions that can generate more power while satisfying the constraints.
In Figs.~\ref{fig:WSvsBP} and \ref{fig:WSvsGT}, the results generally follow the expected trends when compared to the operating point schedule from Fig.~\ref{fig:control profiles}. 
Overall, the optimization-based approach seems to favor larger torque and generator speed values to maximize power production.
As a consequence of relaxing the maximum generator speed from $\omega_{g,\max,1}$ to $\omega_{g,\max,2}$, we see that the optimizer favors lower blade pitch values in the rated region.
The results from the load cases in the below-rated and transition regions are encouraging, as a combination of torque and pitch control is utilized.
In some regions, the pitch control is active, while torque is held constant and vice versa.
Therefore, the optimizer identifies results for all regions in agreement with traditional wind turbine controls.
Overall, these results, in combination with the model validation in Sec.~\ref{sec:IEA15MW}, demonstrate the validity of the considered LPV models in FOWT open-loop control studies.

\subsection{Average Output Power vs.~Plant Design Space}\label{subsec:Pavg-plant}

\begin{figure*}
\centering
\noindent \includegraphics[scale=0.5]{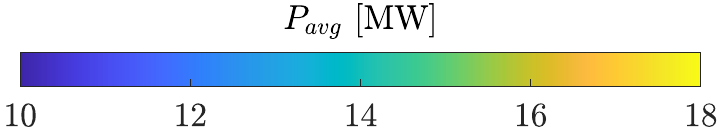}
\begin{subfigure}[b]{0.32\textwidth}
    \centering
    \includegraphics[scale=0.35]{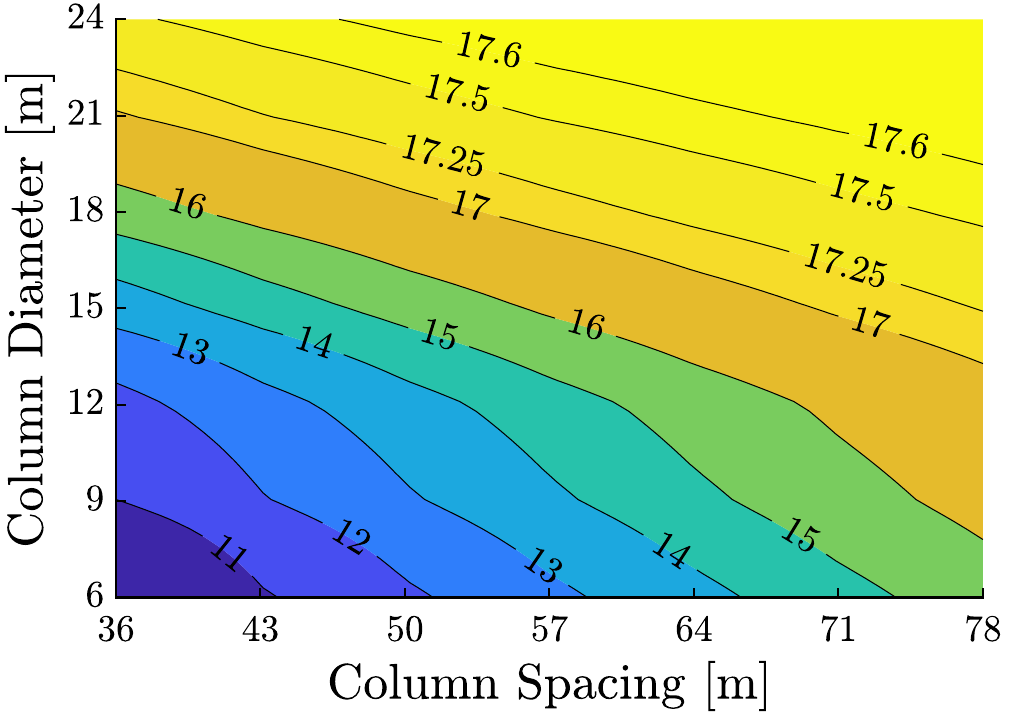}
    \caption{$\Theta_p \leq 3$ [deg].}
    \label{fig:DVvsPwr3}
\end{subfigure}%
%\hspace{0.019\textwidth}%
\begin{subfigure}[b]{0.32\textwidth}
    \centering
    \includegraphics[scale=0.35]{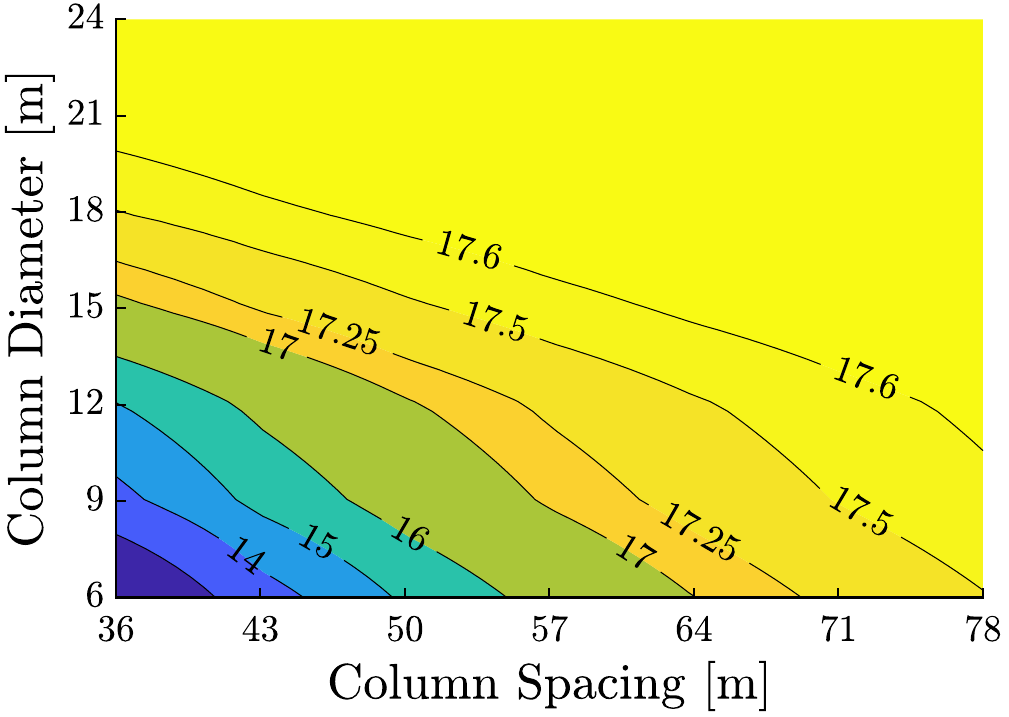}
    \caption{$\Theta_p \leq 5$ [deg].}
    \label{fig:DVvsPwr5}
\end{subfigure}%
%\hspace{0.019\textwidth}%
\begin{subfigure}[b]{0.32\textwidth}
    \centering
    \includegraphics[scale=0.35]{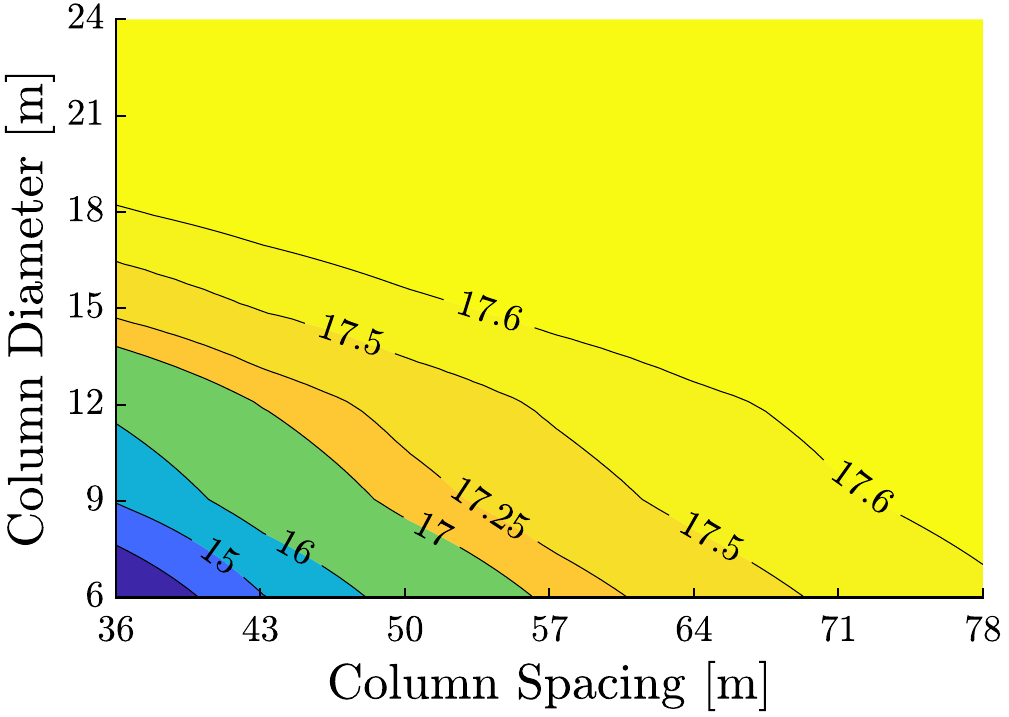}
    \caption{$\Theta_p \leq 6$ [deg].}
    \label{fig:DVvsPwr6}
\end{subfigure}%
\caption{Average power for case 7 with a mean wind speed of $14$ [m/s] vs.~plant design space for different platform pitch ($\Theta_p$)} values.
\label{fig:DVvsPower}
\end{figure*}
%--
%\input{figures/S5/figAEP/DVvsAEP}
\begin{figure*}
\centering
\noindent \includegraphics[scale=0.5]{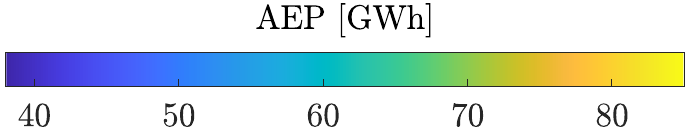}
\begin{subfigure}[b]{0.32\textwidth}
    \centering
    \includegraphics[scale=0.35]{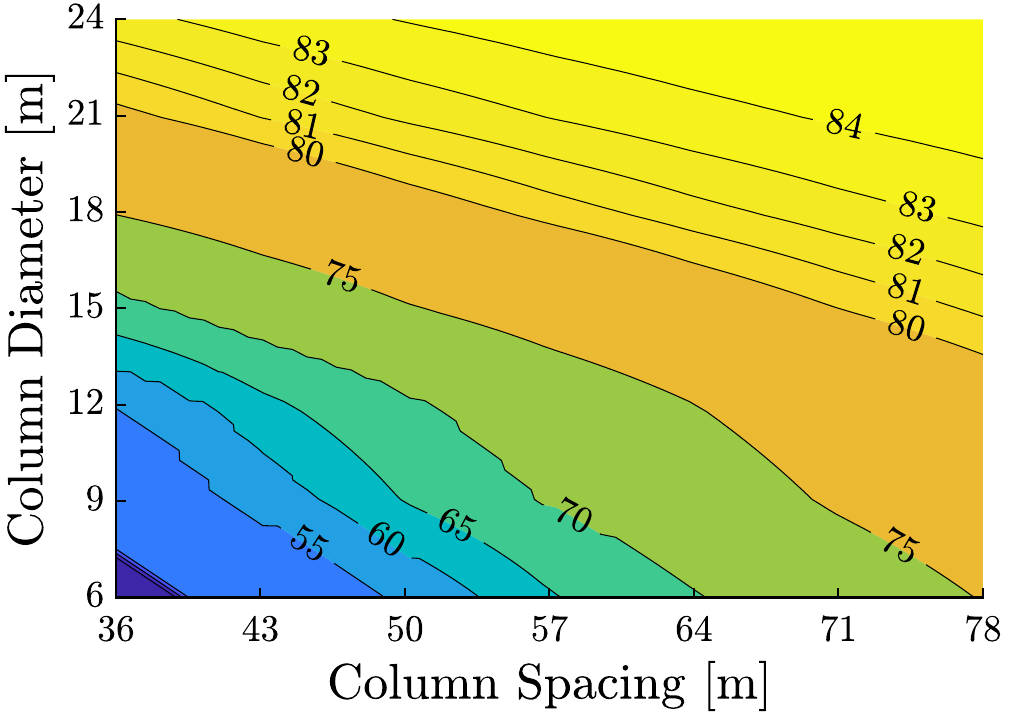}
    \caption{$\Theta_p \leq 3$ [deg].}
    \label{fig:DVvsAEP4}
\end{subfigure}%
%\hspace{0.019\textwidth}%
\begin{subfigure}[b]{0.32\textwidth}
    \centering
    \includegraphics[scale=0.35]{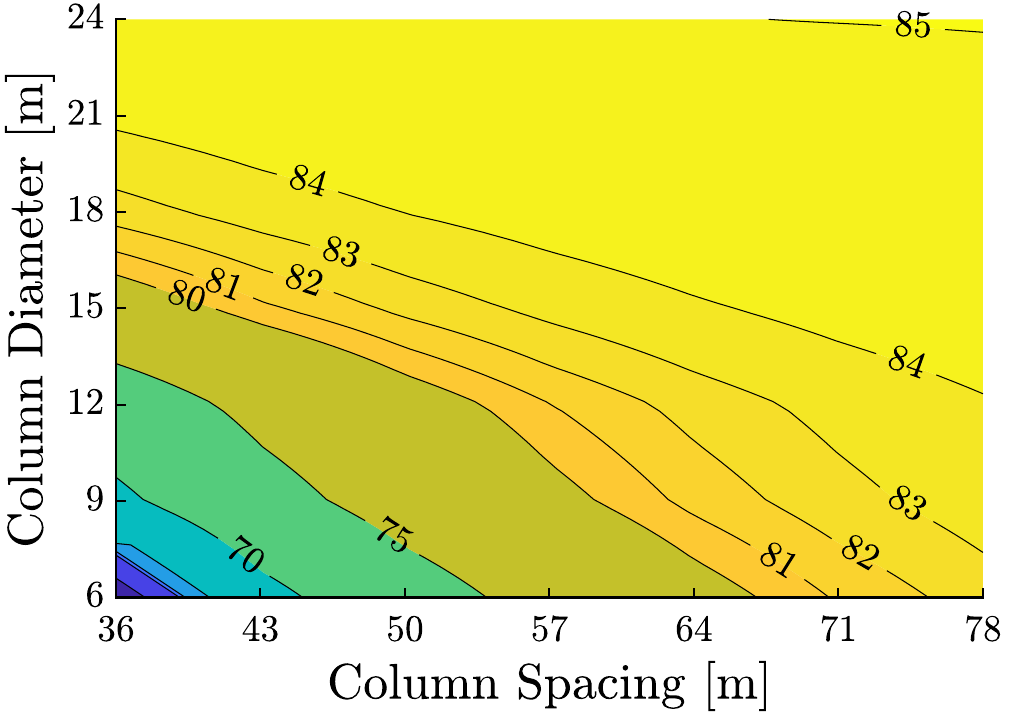}
    \caption{$\Theta_p \leq 5$ [deg].}
    \label{fig:DVvsAEP5}
\end{subfigure}%
%\hspace{0.019\textwidth}%
\begin{subfigure}[b]{0.32\textwidth}
    \centering
    \includegraphics[scale=0.35]{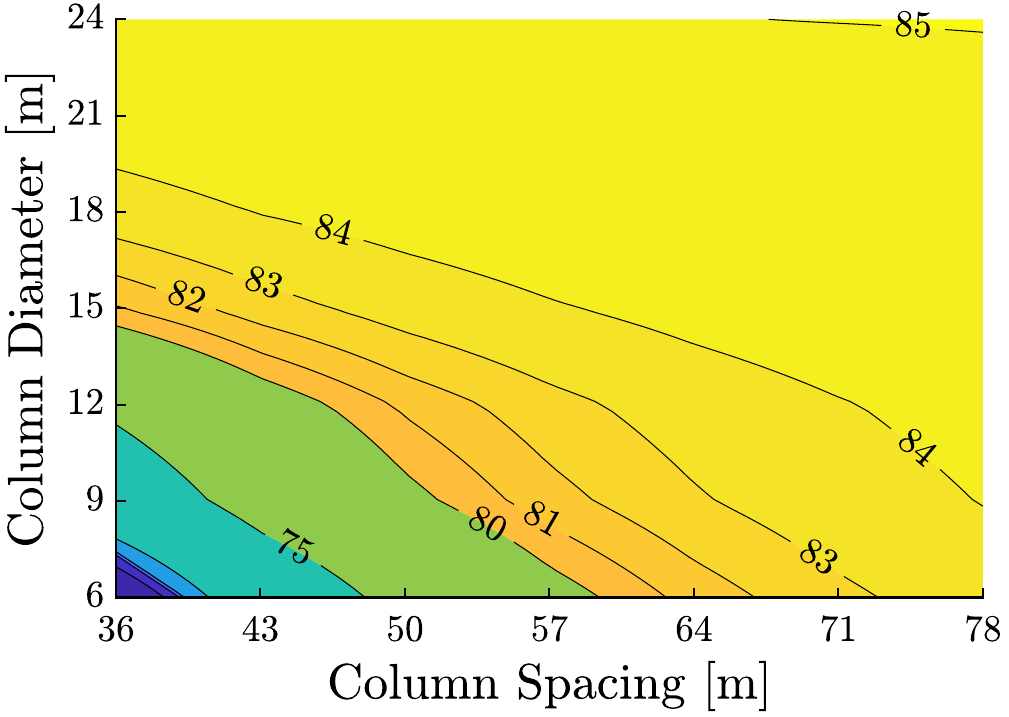}
    \caption{$\Theta_p \leq 6$ [deg].}
    \label{fig:DVvsAEP6}
\end{subfigure}%
\caption{AEP vs.~plant design space for different platform pitch ($\Theta_p$) values.}
\label{fig:DVvsAEP}
\end{figure*}
%--
%\input{figures/S5/figLCOE/LCOEvsDV}
\begin{figure*}
\centering
\noindent \includegraphics[scale=0.5]{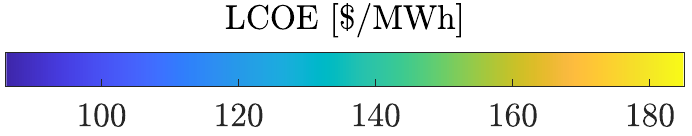}
\begin{subfigure}[b]{0.32\textwidth}
    \centering
    \includegraphics[scale=0.35]{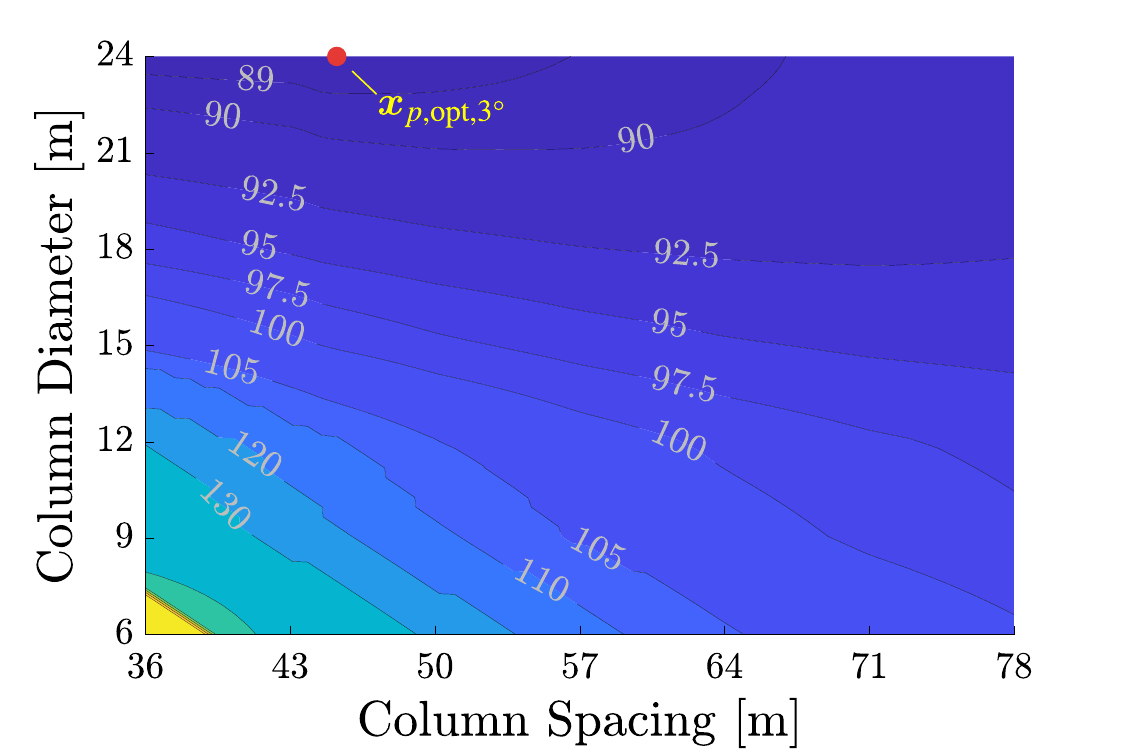}
    \caption{$\Theta_p \leq 3$ [deg].}
    \label{fig:DVvsLCOE4}
\end{subfigure}%
%\hspace{0.019\textwidth}%
\begin{subfigure}[b]{0.32\textwidth}
    \centering
    \includegraphics[scale=0.35]{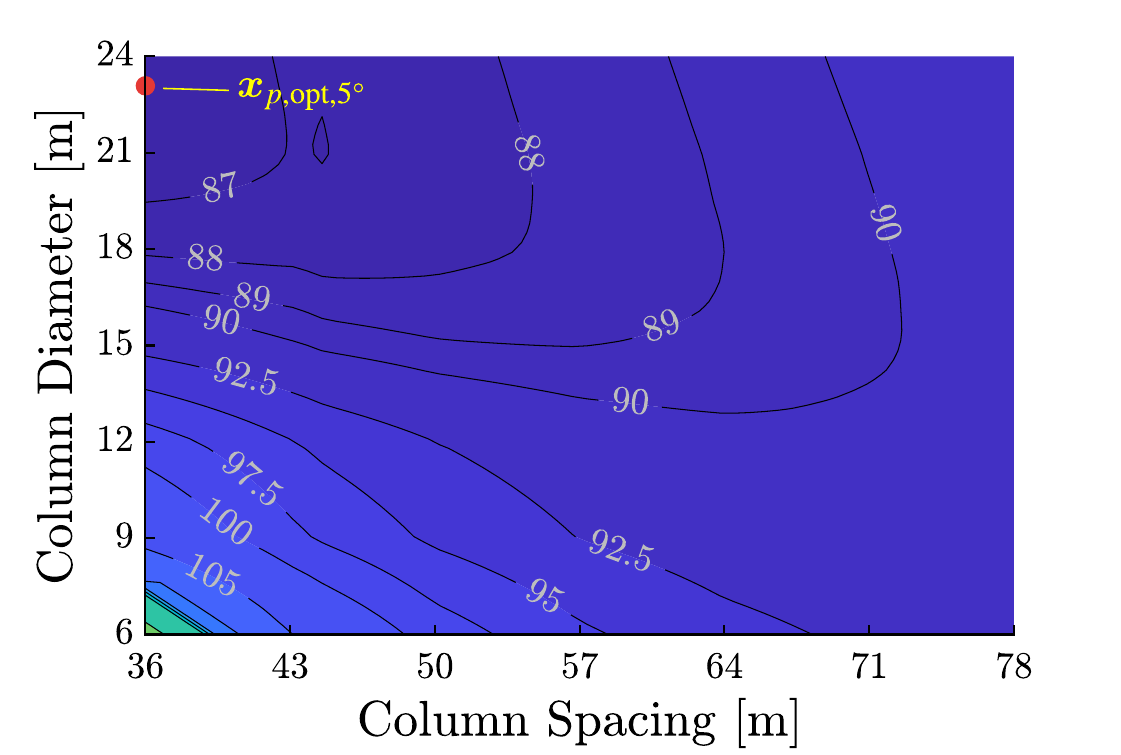}
    \caption{$\Theta_p \leq 5$ [deg].}
    \label{fig:DVvsLCOE5}
\end{subfigure}%
%\hspace{0.019\textwidth}%
\begin{subfigure}[b]{0.32\textwidth}
    \centering
    \includegraphics[scale=0.35]{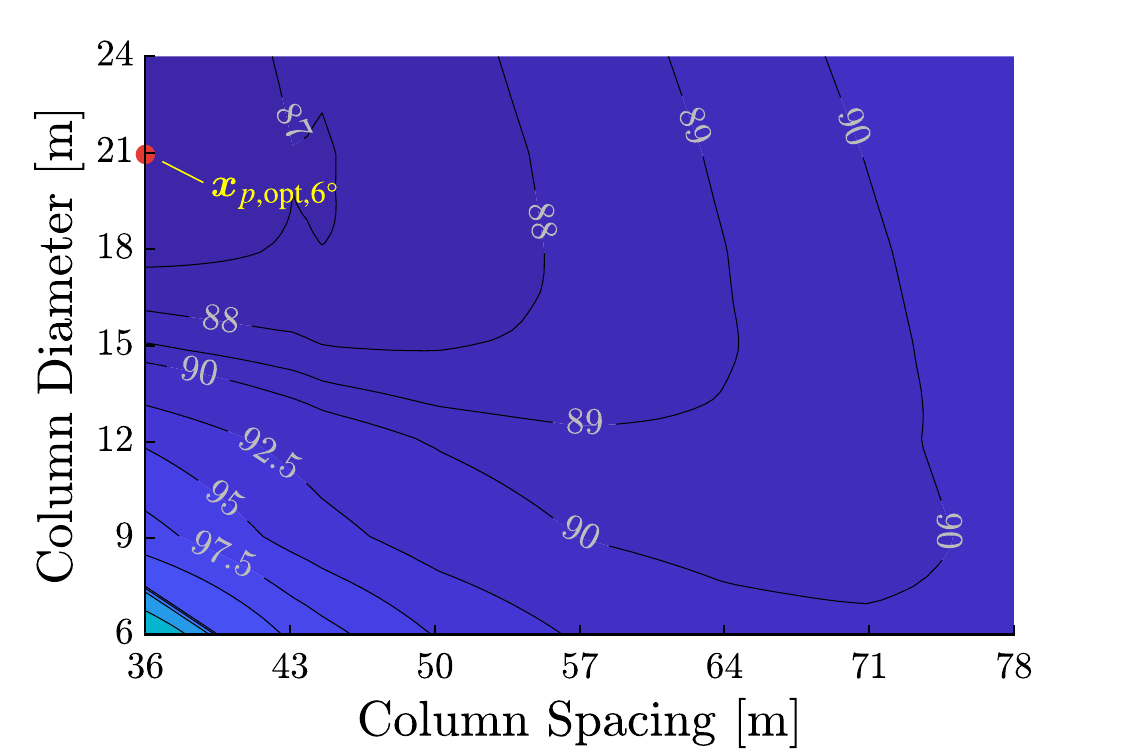}
    \caption{$\Theta_p \leq 6$ [deg].}
    \label{fig:DVvsLCOE6}
\end{subfigure}\\

\caption{LCOE vs.~plant design space  for different platform pitch ($\Theta_p$) values.}
\label{fig:LCOE}
\end{figure*}
%--

In Fig.~\ref{fig:DVvsPower}, the trends between the average power $\bar{P}^*(\bm{x}_p)$ for load case 7 are shown for three of the five tested values of $\Theta_{p,\max}$.
The primary method used to control the platform pitch is the blade pitch $\beta$, but $\beta$ is also tightly coupled to the generator speed.
To satisfy smaller, more challenging values of $\Theta_{p,\max}$, the optimal control solution has higher values of blade pitch, sacrificing generator speed.
Thus, for these more challenging constraint values, the power produced is lower on average.
Additionally, the platform design has a significant effect on the average power production.
Larger values of column spacing $c_s$ and column diameter $c_d$ yield platforms that satisfy the pitch constraints with little to no compromise on power generation. 
In comparison, designs with smaller values of $c_s$ and $c_d$ must sacrifice power generation in some regions.
In addition to these trends for the average output power, we briefly looked at how, for the same DLC, the optimal trajectories of $\tau_g$ and $\beta$ change with $\bm{x}_p$. 
The mean value of $\tau_g$ does not change as $\bm{x}_p$ changes, as the optimizer seeks to maximize the power generated.
This trend holds for all three windspeed regions.
However, for the same DLC, the mean value of $\beta$ is higher for designs with lower values of $c_s$ and $c_d$.
The mean value is disproportionately higher in the transition region as a higher control effort is needed to satisfy the constraint on $\Theta_p$ for these designs.

For some combination of platform pitch constraints and plant design considered in this study, the inner-loop optimizer returns an infeasible result.
These infeasible cases happen primarily for designs with lower values of $c_s$ and $c_d$ and load cases in the transition region, because the system tends to have higher values of platform pitch in this region.
In addition to the cases from the transition region, some load cases in the rated region fail for these plant designs.
For these cases that fail in the rated region, the upper limit on the control blade pitch considered in these studies is insufficient for the optimizer to find feasible solutions.

\subsection{LCOE vs.~Plant Design Space}\label{subsec:LCOE-plant}

% new paragraph
Combining the average power produced for each load case using the scheme in Eq.~(\ref{eq:Eweibull}), we can determine the total energy output.
In addition, utilizing the total cost model mentioned in Sec.~\ref{sec:PlantDesign}, the system LCOE can be estimated.
As mentioned previously, some values of the constraints are infeasible, and the infeasible results are included with zero generated energy.
The summarized LCOE and AEP results are shown in Figs.~\ref{fig:DVvsAEP}  and \ref{fig:LCOE}, respectively.
The Weibull distribution used in the AEP calculation in Eq.~(\ref{eq:Eweibull}) (and shown in Fig.~\ref{fig:DLC_weight}) weights the power produced by the wind cases in the below-rated and transition regions higher than the power produced in the rated region.
Therefore, the transition region will be critical to reducing LCOE, and designs with fewer infeasible cases here would be strongly preferred.

\begin{figure}[t]
\centering 
\noindent \includegraphics[width = 0.4\textwidth]{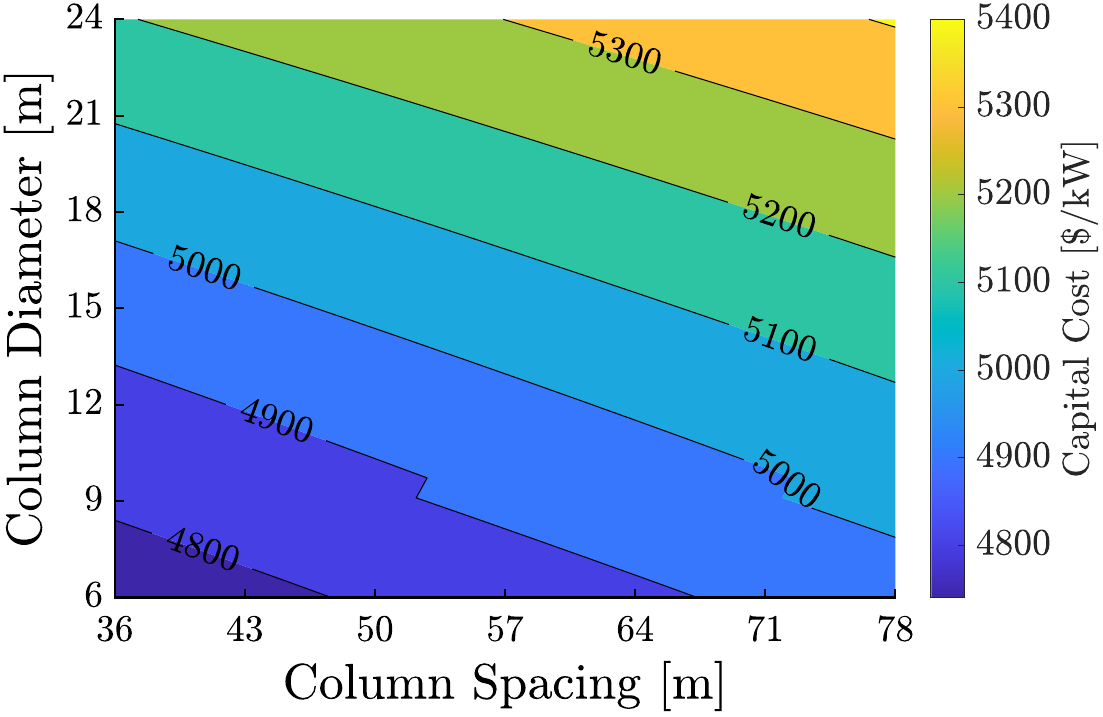}
\caption{Capital cost vs.~plant design space.}
\label{fig:capital_cost}
\end{figure}

% new paragraph
From these results, we see that the optimal value for LCOE depends on the platform pitch constraint. %  \xchange{and the LCOE subspace changes as the constraint values are changed.
The capital cost increases monotonically as $\bm{x}_p$ increases, with a minimum cost of $4,740.7~[\$/\textrm{kW}]$ at $\bm{L}_p = [36,6]^T$, and a maximum of $5,407.2  ~[\$/\textrm{kW}]$ at $\bm{U}_p = [78,24]^T$, as shown in Fig.~\ref{fig:capital_cost}.
Similarly, the AEP increases as $\bm{x}_p$ increases.

For the IEA 15-MW reference turbine described in~\cite{Gaertner2020, Allen2020}, $\Theta_p$ was constrained to $6^\circ$ using the nominal platform dimensions.
From Fig.~\ref{fig:DVvsLCOE6}, we see there is a region that balances the capital cost and power production and, consequently, has lower LCOE values.
While keeping the other plant parameters constant, the design with the lowest LCOE of $86.27$ [\$/\textrm{MWh}] can be obtained using a platform with $\bm{x}_{p,\textrm{opt},6^\circ} = [36.0,20.9]^T$.
Additionally, the lowest LCOE values across all constraints can be found in the neighborhood of this point.
For comparison, the LCOE value for the nominal platform with dimensions $\bm{x}_{p,\textrm{nominal},6^\circ}$ evaluated using this approach is $89.30~[\$/\textrm{MWh}]$.

\begin{figure*}[h]
\centering
\includegraphics[width=0.3\textwidth]{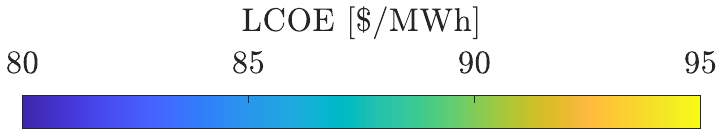}\\
\begin{subfigure}[t]{0.25\textwidth}
\centering
\noindent\includegraphics[scale=0.28]{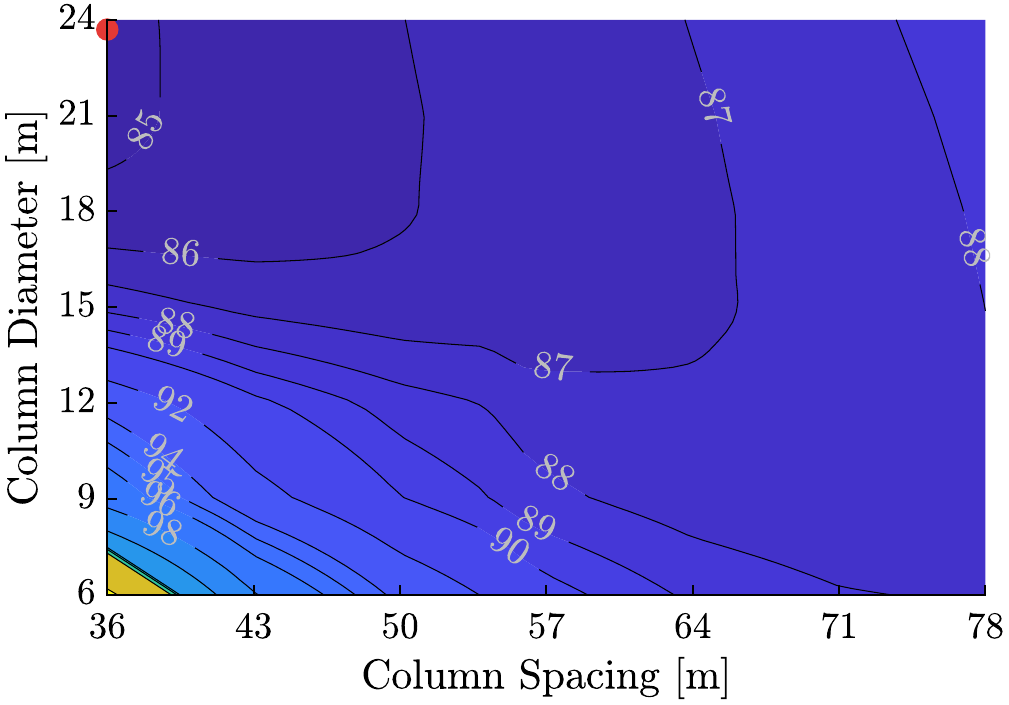}
\caption{$\bm{F} = [0.8,0.8]^T$.}
\label{fig:lcoe_ex1} 
\end{subfigure}%
\begin{subfigure}[t]{0.25\textwidth}
\centering
\noindent\includegraphics[scale=0.28]{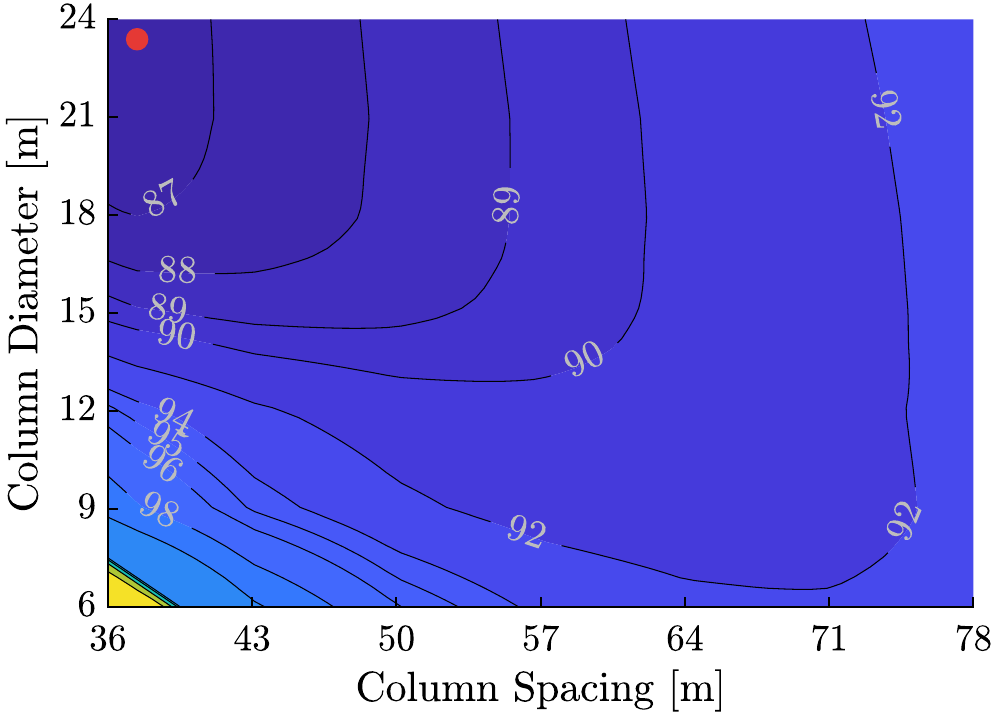}
\caption{$\bm{F} = [1.2,0.8]^T$.}
\label{fig:lcoe_ex2} 
\end{subfigure}%
\begin{subfigure}[t]{0.25\textwidth}
\centering
\noindent\includegraphics[scale=0.28]{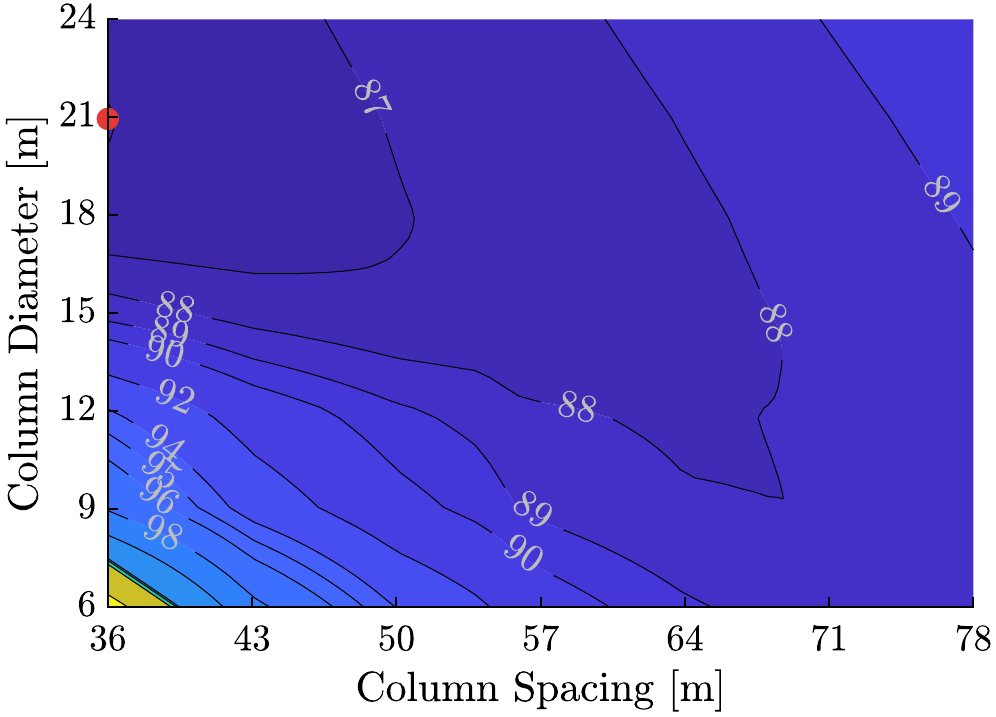}
\caption{$\bm{F} = [0.8,1.2]^T$.}
\label{fig:lcoe_ex3}
\end{subfigure}%
\begin{subfigure}[t]{0.25\textwidth}
\centering
\noindent\includegraphics[scale=0.28]{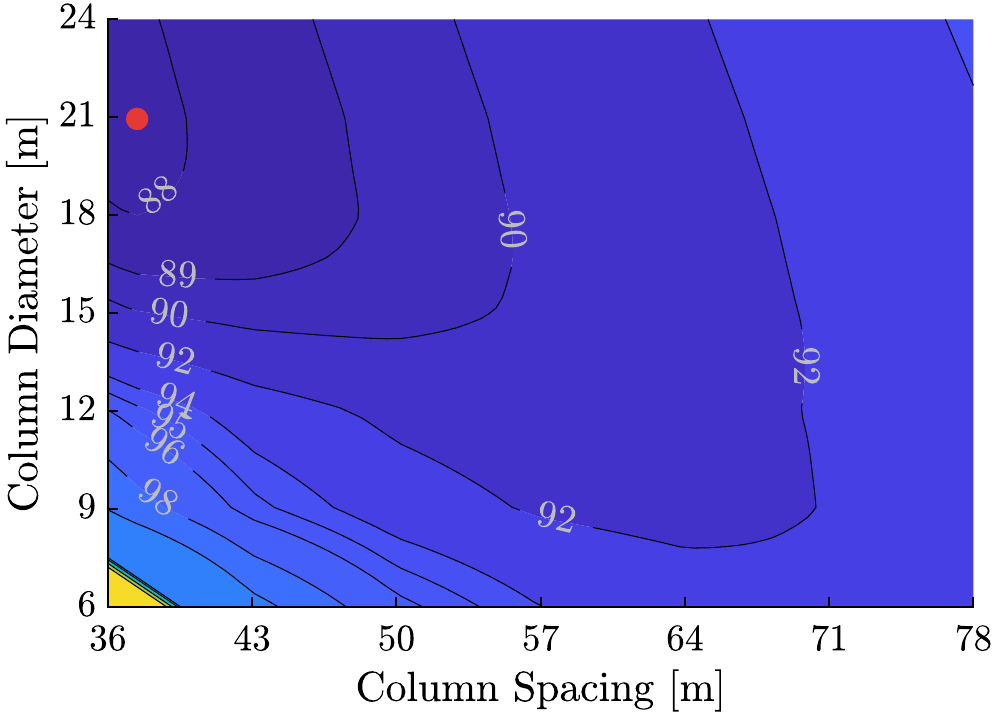}
\caption{$\bm{F} = [1.2,1.2]^T$.}
\label{fig:lcoe_ex4}
\end{subfigure}%
\caption{LCOE vs. plant design space for $\Theta_p \leq 6$ for four different values of the capital cost scaling factor $\bm{F}$. }
\label{fig:LCOE_sens}
\end{figure*}

To explore the sensitivity of the optimization result to variations in the cost model, we consider a variability of $\pm 20 \%$ of the capital cost for both $c_s$ and $c_d$.
Assuming the capital costs of $c_s$ and $c_d$ are independent, the capital cost from Eq.~(\ref{eq:cost}) can be represented as:
\begin{align}
    \label{eq:capital-cost-sep}
    C_{\textrm{capital}}(\bm{x}_p) = C_s(c_s) + C_d(c_d) 
\end{align}
The variations in the cost can be represented through a scaling factor $\bm{F}$ as:
The variations in the cost can be represented through a scaling factor $\bm{F}$ as:
\begin{align}
\label{eq:capital-cost-sens}
    C_{\textrm{capital}}(\bm{x}_p) = \bm{F}^T\begin{bmatrix}C_s(c_s) \\ C_d(c_d)
    \end{bmatrix}^T
\end{align}
\noindent with $\bm{F} =[1,1]^T$ for Eq.~(\ref{eq:capital-cost-sep}).
Figure~\ref{fig:LCOE_sens} shows how the LCOE subspace varies at the four extremities of this uncertainty set.
The optimal LCOE values for these four cases are:
\begin{enumerate}
    \item LCOE = $84.79$ [\$/MWh] at $\bm{x}_p = [36.0,23.6]^T$ for $\bm{F} = [0.8,0.8]^T$, shown in Fig.~\ref{fig:lcoe_ex1}.
    \item LCOE = $86.53$  [\$/MWh] at $\bm{x}_p = [37.4,23.3]^T$ for $\bm{F} = [1.2,0.8]^T$, shown in Fig.~\ref{fig:lcoe_ex2}.
    \item LCOE = $85.97$  [\$/MWh] at $\bm{x}_p = [36.0,20.9]^T$ for $\bm{F} = [0.8,1.2]^T$, shown in Fig.~\ref{fig:lcoe_ex3}.
    \item LCOE = $87.71$ [\$/MWh] at $\bm{x}_p = [37.4,20.9]^T$ for $\bm{F} = [1.2,1.2]^T$, shown in Fig.~\ref{fig:lcoe_ex4}.
\end{enumerate}
Since the AEP does not vary with the cost model, the optimal point is still in the neighborhood of $\bm{x}_{p,\textrm{opt},6^\circ}$, shown in Fig.~\ref{fig:DVvsLCOE6}, as this is the region with maximum AEP and minimum cost.
Because of this, the optimal value of $c_s$ does not change much.
However, the optimal design is more sensitive towards changes in the capital cost associated with $c_d$.
By reducing the cost of $c_d$, the optimum design has a higher value of $c_d$ as shown in Figs.~\ref{fig:lcoe_ex1} and \ref{fig:lcoe_ex2}.

The results presented in this study are subject to modeling assumptions, optimal control operation, and lack of safety factors, but it can still help guide the final design.
Additionally, the hydrodynamic and hydrostatic stability of the different platforms have not been evaluated in this study, along with other DLCs that are meant to test the turbine under fatigue and extreme loading conditions.
These investigations will also limit the bounds on the plant design variables and impact the final design.
%_------------------------------------------------------------
% Conclusion
\xsection{Conclusion}\label{sec:conclusion}

% \subsection{Future Work}
In this work, we discussed the use of LPV models for CCD of FOWTs.
High-fidelity models of FOWTs are described by highly complex and nonlinear models.
Unfortunately, these models are often too costly to use in early-stage system design and evaluation.
Using linearized models based on these nonlinear systems is a popular method to offset the computational costs.
Here, we describe a class of LPV models that realize more accurate predictions of a system's dynamic behavior over a large range of operating points and are shown to be useful for early-stage CCD studies of FOWTs.

The specific FOWT system considered was the IEA 15-MW reference turbine~\cite{Gaertner2020} on a semisubmersible platform~\cite{Allen2020}.
The LPV models based on the wind speed parameter showed good general agreement in both nonlinear simulation comparisons and general optimal control trends. 
The primary study investigated the system's pitching motion as a proxy of its dynamic stability, power production, and, ultimately, the LCOE.
The plant decisions in this study were the distance between the central and outer columns of the platform, along with the diameter of the outer columns, and the results indicated that a system with lower column spacing and higher column diameter values has optimal LCOE values.
The optimal platform design obtained through the proposed approach can satisfy the platform pitch constraints while providing a lower LCOE value.
However, several additional factors should be investigated before making a specific recommendation.

% new paragraph
It remains to future work to incorporate more detailed and sophisticated outer-loop plant design optimization, including the impact of plant decisions, such as tower hub height, blade length, and the mooring system on the platform stability, and power production in the context of the LCOE.
More scalable and efficient strategies for sampling and interpolation must be explored to support the expanded plant model.
Leveraging the LPV model structure for uncertainty propagation in the time domain would support future CCD studies that directly incorporate uncertainties and reliability constraints.
The performance and trade-offs of the LPV approach presented in this article should be compared with approaches for nonlinear derivative function surrogate models~\cite{Deshmukh2017}.
Additionally, we hope to study the effect of wave and current excitations.
Finally, to address the realizability of the open-loop optimal control solutions, work is needed to realize robust, implementable control systems, which may be informed by the optimal operation identified in this study~\cite{Jonkman2021a, Deshmukh2015a}.
%--------------------------------------------------------

%\clearpage
\begin{acknowledgment}
This work was authored in part by the National Renewable Energy Laboratory, operated by Alliance for Sustainable Energy, LLC, for the U.S. Department of Energy (DOE) under Contract No. DE-AC36-08GO28308. Funding provided by the U.S. Department of Energy Office of Energy Efficiency and Renewable Energy Wind Energy Technologies Office. The views expressed in the article do not necessarily represent the views of the DOE or the U.S. Government. The U.S. Government retains and the publisher, by accepting the article for publication, acknowledges that the U.S. Government retains a nonexclusive, paid-up, irrevocable, worldwide license to publish or reproduce the published form of this work, or allow others to do so, for U.S. Government purposes.
The authors would like to thank Alan Wright, Garrett Barter from NREL, Saeed Azad from CSU, and John Jasa from NASA Glenn Research Center for their feedback and suggestions.
\end{acknowledgment}

%--------------------------
% \clearpage
\renewcommand{\refname}{REFERENCES}
\bibliographystyle{asmems4}
\begin{mySmall}

\bibliography{production}
\end{mySmall}

\end{document}